\documentclass[letterpaper,11pt,fleqn]{article}
\pdfoutput=1
\usepackage[margin=1in]{geometry} %
\usepackage[utf8]{inputenc}
\usepackage{amsmath}
\usepackage{amssymb} 
\usepackage{mathtools}
\usepackage{booktabs} 
\usepackage{colortbl} 
\usepackage{dsfont} 
\usepackage[small]{caption} 
\usepackage{subcaption}
\usepackage{booktabs}
\usepackage{mathrsfs}	
\usepackage{cite} 
\usepackage{cancel}
\usepackage{nicefrac}
\usepackage{pifont} 
\usepackage[normalem]{ulem}
\usepackage[pdftex,hidelinks,colorlinks=true,citecolor=magenta,linkcolor=blue,urlcolor=blue]{hyperref}
\usepackage[noabbrev]{cleveref}
\usepackage{hyperref}
\usepackage{braket}
\usepackage{xfrac}
\usepackage{wrapfig}
\usepackage{microtype}
\usepackage[nolist]{acronym}
\usepackage{float}
\usepackage{longtable}
\usepackage{feynmp-auto}
\usepackage{amsfonts,stackengine,graphicx}
\usepackage{mwe}
\usepackage{bbm}
\usepackage{xspace}
\usepackage{enumerate} 
\usepackage{pdflscape}
\usepackage{multirow}
\usepackage{array}
\newcolumntype{L}{>{$}l<{$}} 
\usepackage{comment}
\usepackage{color}
 \usepackage{hyperref}

\newcolumntype{C}{>$c<$}
\usepackage{bbold}

\usepackage[all,cmtip]{xy}
\newlength{\xywd}
\newcommand{\xyrightarrow}[2][]{%
  \sbox{0}{$\scriptstyle#1$}%
  \xywd=\wd0
  \sbox{0}{$\scriptstyle#2$}%
  \ifdim\wd0>\xywd \xywd=\wd0 \fi
  \xymatrix@C\dimexpr\xywd+1em\relax{{}\ar[r]^{#2}_{#1}&{}}%
}

\usepackage[textsize=tiny,backgroundcolor=yellow,textwidth=80pt]{todonotes}

\DeclareMathOperator{\re}{Re}

\newcommand{\SU}[1]{\ensuremath{\mathrm{SU}(#1)}}
\newcommand{\U}[1]{\ensuremath{\mathrm{U}(#1)}}

\newcommand{\FN}{\ensuremath{\mathrm{FN}}}
\newcommand{\fl}{\ensuremath{\mathrm{flat}}}
\newcommand{\pl}{\ensuremath{\mathrm{pl}}}
\newcommand{\CP}{\ensuremath{CP}\xspace}

\usepackage{xcolor}

\newcommand{\dd}{\mathop{}\!\mathrm{d}}
\newcommand{\ii}{\hskip0.1ex\mathrm{i}\hskip0.1ex}
\newcommand{\ee}{\mathrm{e}}
\definecolor{darkgreen}{HTML}{109930}
\definecolor{darkbrown}{HTML}{8B4513}
\definecolor{pink}{rgb}{0.858, 0.188, 0.478}
\usepackage[textsize=tiny]{todonotes}

\begin{document}

\newcommand\mytitle{INFLAVON\\[1em] 
\it{CMB as cosmic tracer of Flavor physics}} 

\begin{titlepage}
\begin{flushright}
UCI-TR-2026-02
\end{flushright}

\vspace*{2em}

\begin{center}
{\LARGE\sffamily\bfseries\mytitle}

\vspace{2em}

\renewcommand*{\thefootnote}{\fnsymbol{footnote}}

\textbf{%
Mu--Chun Chen$^{a}$\footnote{muchunc@uci.edu},
Anish Ghoshal$^{b}$\footnote{anish.ghoshal@fuw.edu.pl}, 
V.~Knapp--P\'erez$^{a}$\footnote{vknapppe@uci.edu},
Xueqi Li$^{a}$\footnote{xueqi.li@uci.edu},
Xiang--Gan Liu$^{a}$\footnote{xianggal@uci.edu},
Cameron Moffett-Smith$^{a}$\footnote{cmmoffet@uci.edu}}
\\[8mm]
\textit{$^a$\small
~Department of Physics and Astronomy, University of California, Irvine, CA 92697-4575 USA}
\\[5mm]
\textit{$^b$\small Department of Physics and Astronomy, University of Sussex, \\
Brighton,  BN1 9RH, United Kingdom}
\end{center}

\vspace*{5mm}

\begin{abstract}

We unify one of the most widely studied frameworks to explain the hierarchical structure of the flavor sector in the \ac{SM}, the \ac{FN} mechanism, with cosmic inflation. We propose that the complex scalar field, the so-called \textit{flavon}, which breaks the \ac{FN} $\U{1}$ symmetry and generates the Yukawa couplings of the $\ac{SM}$, to also drive inflation, which we dub as \textit{Inflavon}. After inflation ends, the decay of the \textit{inflavon} reheats the Universe, establishing a novel link between early Universe cosmology and flavor physics. As concrete examples, we present realizations where the \textit{inflavon} potential is described by an $\alpha-$attractor potential. We then compute the resulting \ac{CMB} observables, specifically the spectral index ($n_s$), the tensor-to-scalar ratio ($r$), and the amplitude of scalar perturbations ($A_s$) as functions of the underlying \ac{FN} model parameters.
We identify the parameter space in \ac{FN} models involving the scale of Flavor symmetry breaking $\Lambda_{\rm FN}$ and FN charges which is ruled out by Planck and ACT data, as well as the region that could be probed by next-generation \ac{CMB} experiments like CMB-S4, SO and LiteBIRD. We also discuss $inflavon$ as dark matter and its isocurvature constraints.
\end{abstract}

\end{titlepage}
\renewcommand*{\thefootnote}{\arabic{footnote}}
\setcounter{footnote}{0}


\section{Introduction}
\label{sec:Intro}

The Standard Model (SM) of particle physics is very successful in depicting all the known fundamental particles and interactions to great accuracy. However, the \ac{SM} does not explain the existence of three families of fermions and the pattern of their masses and mixings. For instance, in the quark sector, the mixing between families is small, as encoded in the parameters of the \ac{CKM} matrix. In contrast, the lepton sector exhibits large mixing, as expressed by the parameters of the \ac{PMNS} matrix~\cite{ParticleDataGroup:2024cfk}. This is referred to in the literature as the ``flavor puzzle''.

Among solutions that have been proposed to explain the flavor puzzle, see e.g.~\cite{Feruglio:2015jfa,Xing:2020ijf}, one often invokes additional symmetries beyond those that exist in the \ac{SM}. Symmetries that have been considered are either global or gauged. For example, the well-known Froggatt-Nielsen (FN) mechanism~\cite{Froggatt:1978nt,Leurer:1992wg,Leurer:1993gy} utilizes, most commonly, a $\U{1}_{\FN}$ symmetry whose spontaneous breaking gives rise to the Yukawa couplings as higher-dimensional operators in the \ac{EFT}. This mechanism involves the \ac{SM} fermions, the Higgs field, and the \ac{VEV} of one or more scalar fields responsible for the $\U{1}_{\FN}$ breaking. The original and variations of the \ac{FN} scenarios have been studied for their fingerprints in collider experiments or searches for rare processes~\cite{Tsumura:2009yf,Calibbi:2012at,Bauer:2016rxs,Ema:2016ops,Calibbi:2016hwq,Smolkovic:2019jow}. However, the \ac{FN} scenarios traditionally involve new physics at very high energy scales, beyond the reach of particle physics laboratory experiments.

On the other hand, the paradigm of cosmic inflation is a well-established framework. It can generate the primordial density fluctuations which are responsible for the generation of the large-scale structures as seen in the universe. It also addresses the flatness and the horizon problems of the universe~\cite{Brout:1977ix,Sato:1980yn,Guth:1980zm,Linde:1981mu,Starobinsky:1982ee}. In its simplest form, inflation is driven by a single slowly rolling scalar field with an approximately very flat potential that dominated the energy density in early Universe. All pre-existing matter and radiation present before the inflationary period gets diluted  due to the exponential expansion of the Universe. Therefore, a {\it reheating} mechanism is required to eventually repopulate the Universe with radiation, as indicated by late-time observations. This reheating phenomenon can be realized through some interactions between the inflaton and the \ac{SM} fields \cite{Kofman:1994rk,Kofman:1997yn}, see Ref. \cite{Allahverdi:2010xz} for a review. However the reheating scale, often depicted by the temperature during reheating $T_{\mathrm{RH}}$, is bounded by several constraints. The lower bound is set by successful \ac{BBN} scale, $T_{\mathrm{RH}} \geq 4 $ MeV~\cite{Ichikawa:2005vw, Kawasaki:2000en, Barbieri:2025moq}. The upper bound should respect the maximum allowed energy density of the universe after the inflation era $\rho_{\rm max}\sim 10^{66} \text{ GeV}^4$~\cite{Baumann:2009ds}, as suggested by Planck data~\cite{Planck:2018vyg} which translates into $T_{\rm RH}$ bounds approximately around $10^{16}$ GeV depending upon the details of the reheating models.

For any given inflationary model, the key predictions are the scalar spectral index $n_s$, the tensor-to-scalar ratio $r$,  and the amplitude of scalar perturbations $A_s$. These three quantities depend very intricately on the details of the reheating mechanism. In particular, 
they depend on the number of e-folds from the end of inflation to the beginning of a radiation-dominated Universe. This evolution is tracked via the evolution of the \ac{FRW}  scale factor $a(t)$~\cite{Cook:2015vqa, Ueno:2016dim, Drewes:2017fmn, Maity:2018dgy, Maity:2018exj, Drewes:2019rxn, Haque:2020zco, DiMarco:2021xzk}. From the current cosmological data, $n_s$, $r$ and $A_s$ are tightly constrained. The value of $A_s$ is extracted to be $\left( 2.090 \pm 0.101\right)\times 10^{-9}$ by the Planck 2018 data~\cite{Planck:2018vyg}. Additionally, Planck 2018 data~\cite{Planck:2018jri} also gives the value $n_s = 0.965 \pm 0.004$ at 95 \% \ac{CL} while recent measurements from the Atacama Cosmology Telescope (ACT) collaboration~\cite{ACT:2025tim, ACT:2025fju}, reports $n_s = 0.9666 \pm 0.0077$. Interestingly, once both ACT and Planck datasets are combined, the resulting value of $n_s = 0.9709 \pm 0.0038$ is obtained due to the opposite correlations between $n_s$ and the baryonic abundance $\Omega_b  h^2$. 
Given the shift of $n_s$ towards larger values, the implications and interpretations in the context of inflationary models have led to several recent studies \cite{Kallosh:2025rni, Pan:2025psn, Dioguardi:2025vci, Brahma:2025dio, Gialamas:2025kef, Antoniadis:2025pfa, Gao:2025onc, Wang:2025zri, Yin:2025rrs, Liu:2025qca, Gialamas:2025ofz, McDonald:2025odl}. The
predicted upper limits on $r$ at 95 \% \ac{CL} from Planck and Planck $+$ ACT are
0.036 and 0.038 respectively.

In this paper, we propose to use the data from precision cosmology observation to test the flavor models based on the \ac{FN} mechanism. Specifically, we constructed a unified framework where the flavon field, which is responsible for breaking the $\U{1}_{\FN}$ symmetry to generate the fermion masses and mixing pattern, also plays the role as the inflaton, as originally proposed in~\cite{Ema:2016ops}. However, unlike in~\cite{Ema:2016ops}, the dominant decay channel in our case proceeds not through the right-handed neutrino Majorana mass term, but rather via the  \ac{SM} Yukawa couplings.  We refer to this particle as the {\it inflavon}. This dual role opens up the possibility of tracing the fingerprints of the \ac{FN} mechanism and flavor mixing patterns in \ac{CMB} experiments. Within the perturbative reheating framework, we will show that the flavor mixing pattern is dependent on the reheating history of the Universe. This is because \ac{SM} Yukawa couplings determine the decay widths of the {\it inflavon} during reheating. This provides a pathway to probe \ac{FN} models via \ac{CMB} observables $A_s$, $n_s$ and $r$, that are influenced by physics in the post-inflationary reheating era, in general~\cite{Drewes:2015coa,Drewes:2017fmn,Drewes:2019rxn,Drewes:2022nhu}, and in the context of matter-antimatter asymmetry ~\cite{Ghoshal:2022fud} and of dark matter physics~\cite{Ghoshal:2025ejg}.  

In our framework, the choice of inflationary potential is left unconstrained. However, for definiteness, we choose to work with $\alpha-$attractor potentials. Since the inflavon decay width is primarily influenced by \ac{SM} Yukawa couplings, the reheating temperature is correlated with the flavor mixing parameters.
This, in turn, gives rise to distinct predictions of different \ac{FN} flavor models on the $n_s$-$r$ plane. These predictions are contrasted with the results of the combined data from \textit{Planck}, \textit{BICEP/Keck}, and \textit{BAO}~\cite{BICEP:2021xfz}, and from the combined data of \textit{Planck}, \textit{ACT}, and \textit{DESI}~\cite{ACT:2025tim}. Moreover, we also compare our predictions with the forecasts for the next generation  of \ac{CMB} experiments like~\textit{LiteBIRD}, \textit{CMB-S4}, and \textit{SO}\cite{LiteBIRD:2022cnt,CMB-S4:2016ple,SimonsObservatory:2018koc}.\footnote{During the preparation of this work, the project \textit{CMB-S4} was still ongoing. Even though its status has changed, we keep it in our analysis to contrast with other ongoing/upcoming experiments.} We investigate the sensitivity of the predicted \ac{CMB} observables on the \ac{FN} charge assignments. In particular, we study the capability of \ac{CMB} data for ruling out different \ac{FN} models for a given choice of $\alpha-$attractor potential. Furthermore, we also relate the \ac{FN} scale to inflationary observables, and thus affording a way to probe the \ac{FN} scale. To the best of our knowledge, the present study is perhaps the first attempt to propose such an alternative test via \ac{CMB} precision cosmological measurements to probe the \ac{FN} scale of flavon models, which is generally inaccessible to laboratory experiments.
 
\textit{The paper is organized as follows.} In \Cref{sec:FnModels}, we review the basic framework of \ac{FN} models, show how the flavon field can play the role of an inflaton, and derive an $\alpha$-attractor potential from a non-canonical flavon kinetic term. In \Cref{sec:CMB-alpha}, we briefly revisit how the spectral index $n_s$, the scalar-to-tensor ratio $r$, and the amplitude of scalar perturbations $A_s$ can be related to decay width of the inflaton, following closely~\cite{Drewes:2019rxn}. In \Cref{sec:FNFaceCosmology}, we present our main results. We show the predictions for \ac{CMB} observables in different \ac{FN} models and compare them with experimental observations and forecasts. Finally in  \Cref{sec:Conclusion}, we conclude and discuss potential future directions. Technical details are provided in the Appendices.

\section{$\U{1}$ Froggatt-Nielsen models}
\label{sec:FnModels}
The Froggatt-Nielsen mechanism~\cite{Froggatt:1978nt} provides a compelling framework for addressing the mass hierarchy and flavor mixing for quarks and leptons in the \ac{SM}. In this work, we consider $\U{1}_{\FN}$ \ac{FN} models that extend the \ac{SM} with a complex scalar field $\phi$, the so-called flavon. Furthermore, we introduce  $3$ right-handed neutrinos $N_i$ and assume that neutrino masses are generated via the type-I seesaw mechanism. Without loss of generality, the flavon $\phi$ and the \ac{SM} Higgs $H$ are assumed to have \ac{FN} charge of $-1$ and $0$, respectively. The Lagrangian before $\U{1}_{\FN}$ symmetry breaking is given by  

\begin{align}
    -\mathcal L_{\FN}  &~=~ \alpha_{ij}^u \overline{Q}_i \widetilde{H} u_j \left( \frac{\phi}{\Lambda_\FN } \right)^{|n_{ij}^{u}|} + \alpha_{ij}^d \overline{Q}_i H  d_j \left( \frac{\phi}{\Lambda_\FN } \right)^{|n_{ij}^{d}|}   \nonumber\\
     &~+~ \alpha_{ij}^e  \overline{L}_i H E_j  \left( \frac{\phi}{\Lambda_\FN } \right)^{|n_{ij}^{e}|} + \alpha_{ij}^D  \overline{L}_i \widetilde{H} N_j \left( \frac{\phi}{\Lambda_\FN } \right)^{|n_{ij}^{D}|} \nonumber\\
     &~+~ \alpha_{ij}^N \Lambda_\nu \overline{N^c}_i N_j + \text{h.c.}\;,
    \label{eq:BeforeFlavorL}
\end{align}
where $i, \, j = 1, \, 2, \, 3$ are the generation indices, $\alpha_{ij}^{a}$ for $a=u,d,e,D,N$  are order $\mathcal{O}(1)$ dimensionless coupling constants, $Q_i$ are the $\SU{2}_L$ quark doublets, $u_i$, $d_i$ are the up and down quark $\SU{2}_L$ singlets, $L_i$ are the lepton $\SU{2}_L$ doublets, $\Lambda_\nu$ is the right-handed Majorana neutrino mass scale and $\Lambda_\FN$ is the $\U{1}_{\FN}$ symmetry breaking scale. The exponents $n_{ij}^{a}$ shape the flavor structure and are determined by the \ac{FN} charge $(q)$ assignments via
\begin{align}
    n_{ij}^{u}  & ~=~ q_{Q_{i}} -  q_{u_{j}}\;, \nonumber \\
    n_{ij}^{d}  & ~=~ q_{Q_{i}} -  q_{d_{j}}\;, \nonumber \\
    n_{ij}^{e}  & ~=~ q_{L_{j}} - q_{E_{i}}\;, \nonumber \\
    n_{ij}^{D}  & ~=~   q_{L_{i}}\; . 
    \label{eq:nij}
\end{align}
It is to be understood that when $n_{ij} < 0$, $\phi$ is replaced by $\phi^{\dagger}$ in \cref{eq:BeforeFlavorL}. The right-handed neutrinos are assumed to be neutral under $U(1)_{FN}$. Expanding the flavon field around its \ac{VEV} $\braket{\phi} = v_\phi$, we have
\begin{equation}
\label{eq:flavonVev}
    \phi ~\xmapsto{}~ v_\phi + \sigma + \ii\rho\;,
\end{equation}
where $\sigma$ and $\rho$ are real scalar fields. Substituting \cref{eq:flavonVev} in \cref{eq:BeforeFlavorL} yields the effective Yukawa Lagrangian
\begin{equation}
\label{eq:SM}
    -\mathcal{L}_{\text{Yukawa}} ~=~ Y_{ij}^u \overline Q_i \widetilde H u_j + Y_{ij}^d \overline Q_i H d_j + Y_{ij}^e \overline L_i H E_j + Y_{ij}^D  \overline L_i \widetilde H N_j + \alpha_{ij}^{N} \Lambda_\nu  \overline{N^c}_i N_j + \mathrm{h.c.} \; ,
\end{equation}
where the Yukawa couplings are given by
\begin{equation}
\label{eq:YukawaCouplings}
    Y_{ij}^{a} := \alpha_{ij}^{a} \lambda^{|n_{ij}^{a}|}\; ,
\end{equation}
where we define $\lambda := \frac{v_\phi}{\Lambda_\FN}$. In our analysis, we assume $\lambda = 0.17$. Hence, the flavor structures in the quark and lepton sectors are governed by the set of $\U{1}_\FN$ charges and the $\mathcal{O}\left(1\right)$ dimensionless couplings $\alpha_{ij}^{a}$. 
The dynamics of the flavon field is described by the following non-linear sigma model Lagrangian~\cite{Ema:2016ops,Yi:2021xhw}
\begin{equation}
\label{eq:FlavonPotentialK}
    \mathcal{L}_{\phi} ~=~ K(\phi)|\partial \phi |^2 - \kappa(|\phi|^2 -v_\phi^2)^{2n}\;,
\end{equation}
where $n$ is an integer and $\kappa$ is a parameter with mass-dimension depending on $n$. The function $K(\phi)$ in the non-canonical kinetic term depends on the absolute value $|\phi |$, a dimensionless positive parameter $\alpha$ and the scale at which the inflation potential becomes flat $\Lambda_{\fl}$. In this work, we focus on three different inflationary $\alpha-$attractor potentials \cite{Kallosh:2013lkr,Kallosh:2013hoa,Kallosh:2013yoa,Kallosh:2013pby,Kallosh:2013maa,Galante:2014ifa,Kallosh:2022feu}: the E-model, the T-model and the P-model (or polynomial model). These are obtained from \cref{eq:FlavonPotentialK} and from the $K(\phi)$ functions given by
\begin{equation}
\label{eq:Kahler}
    K(\phi)  ~:=~ \begin{cases}
        \frac{3\alpha}{2}\frac{2 \Lambda_\fl^2}{|\phi|^2}\;, \quad \text{E-Model}\;,\\[2em]
        \frac{3\alpha}{2}\frac{2 \Lambda_\fl^2}{|\phi|^2}\left(\frac{|\phi|^2}{v_\phi^2}-2\right)^{-2}\;,\quad\text{T-Model}\;,\\[2em]
        \frac{3\alpha}{2}\frac{2\Lambda_{\fl}^2}{v_\phi^2}\left(1-\left(1-\frac{|\phi|^2}{v_\phi^2}\right)^2\right)^{-\frac{1}{2n}}\times\left(1-\left(1-\frac{|\phi|^2}{v_\phi^2}\right)^{2n}\right)^{-2}\\
        \times \left[\left(1-\left(1-\frac{|\phi|^2}{v_\phi^2}\right)^{-2n}\right)^{-\frac{1}{2n}}+\left(\frac{|\phi|^2}{v_{\phi}^2}\right)^{-\frac{1}{2n}}\right]\;,\quad\text{P-Model}\;.
    \end{cases}
\end{equation}
We then can define the canonically normalized field $\chi$ through the following relations
\begin{equation}
\label{eq:ChiAndTheta}
    \phi~:=~ \begin{cases}
        v_\phi\exp\left[ - \sqrt{\frac{2}{3\alpha}}\frac{\chi+\ii\theta}{2\Lambda_\fl}\right]\;,\quad\text{E-Model}\;,\\[2em]
        v_\phi \sqrt{1+\tanh\left(\sqrt{\frac{2}{3\alpha}}\frac{\chi}{\Lambda_{\fl}}\right)}\exp\left[ - \ii\sqrt{\frac{2}{3\alpha}}\frac{\theta}{2\Lambda_{\fl}}\right]\;,\quad\text{T-Model}\;,\\[2em]
        v_\phi\sqrt{1+\left( \frac{\chi^{2n}}{\chi^{2n}+\left( \sqrt{\frac{3\alpha}{2}}\Lambda_{\fl}\right)^{2n}}\right)^{\frac{1}{2n}}}\exp\left[ - \ii\frac{\theta}{\sqrt{6\alpha}\Lambda_{\fl}}\right]\;,\quad\text{P-Model}\;.
    \end{cases}
\end{equation}
Here $\theta$ is the axion field whose normalization, in general, depends on the \ac{VEV} $v_\phi$. Then, the potential from \cref{eq:FlavonPotentialK}, after the field redefinition of \cref{eq:ChiAndTheta}, is given by
\begin{equation}
    V\left(\chi,\alpha, n,\Lambda_\mathrm{inf},\Lambda_\fl\right) ~:=~ \begin{cases}
        \Lambda_{\mathrm{inf}}^4\left(1-\exp\left[-\sqrt{\frac{2}{3\alpha}}\frac{\chi}{\Lambda_{\fl}}{}\right]\right)^{2n}\;,\quad\text{E-Model}\;,\\[2em]
        \Lambda_{\mathrm{inf}}^4\left(\tanh\left[ \sqrt{\frac{2}{3\alpha}}\frac{\chi}{\Lambda_\fl}\right]\right)^{2n}\;,\quad\text{T-Model}\;,  \\[2em]
         \Lambda_{\mathrm{inf}}^4\frac{\chi^{2n}}{\chi^{2n}+\left( \sqrt{\frac{3\alpha}{2}}\Lambda_\fl\right)^{2n}}\;,\quad\text{P-Model}\;.
    \end{cases}
    \label{eq:alpha-attractors}
\end{equation}
where $\chi$ is the inflaton, and $\Lambda_{\text{inf}}^4 = \kappa v_\phi^{4n}$ denotes the characteristic energy scale in the $\alpha$-attractor potential. As we will show later, this parameter is determined by requiring consistency with the observed amplitude of scalar perturbations, $A_s$. Consequently, $\Lambda_{\inf}$ and $\Lambda_{\fl}$ need not coincide.
These potentials are very well motivated in several cosmological scenarios. Some well known cases are the Starobinsky inflation ($\alpha=1$) \cite{Starobinsky:1980te}, the Goncharov-Linde model ($\alpha=1/9$) \cite{Goncharov:1984jlb,Linde:2014hfa} and the Higgs Inflation scenario ($\alpha=\sqrt{2/3}$) \cite{Bezrukov:2007ep}.  More generally, these so called $\alpha$-attractor models have been shown to form universality class \cite{Kallosh:2013hoa}. Several superstring-inspired scenarios, like K\"ahler Moduli Inflation, Poly-instanton Inflation and Fiber Inflation fall within this framework with  \mbox{$3\alpha=1,2,3,4,5,6,7$} \cite{Ferrara:2016fwe,Kallosh:2017ced} (e.g., Fiber Inflation with $\alpha=2$ and 1/2 \cite{Cicoli:2008gp,Kallosh:2017wku}).
Alternatively, in no-scale supergravity, arbitrary values of $\alpha$, including both \mbox{$\alpha < 1$} and \mbox{$\alpha > 1$}, can be accommodated~\cite{Ellis:2019bmm,Ellis:2019hps}.

\begin{figure}[t]
    \centering
    \begin{subfigure}[t]{0.485\linewidth}
        \centering
        \includegraphics[width=\linewidth]{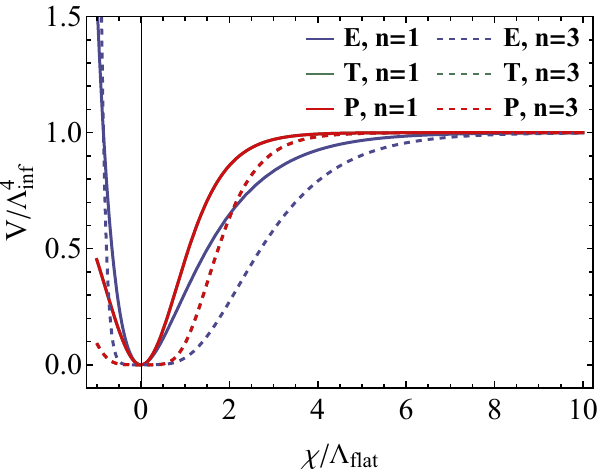}
        \caption{$\alpha = 1$}
    \end{subfigure}
    ~
    \begin{subfigure}[t]{0.485\linewidth}
        \centering
        \includegraphics[width=\linewidth]{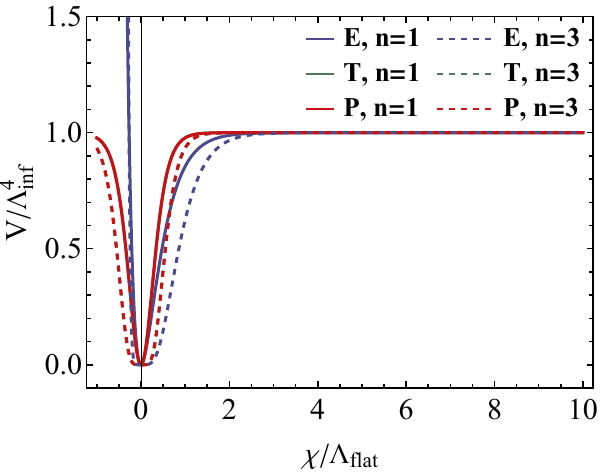}
        \caption{$\alpha = 0.1$}
    \end{subfigure}
    \\
    \begin{subfigure}[t]{0.485\linewidth}
        \centering
        \includegraphics[width=\linewidth]{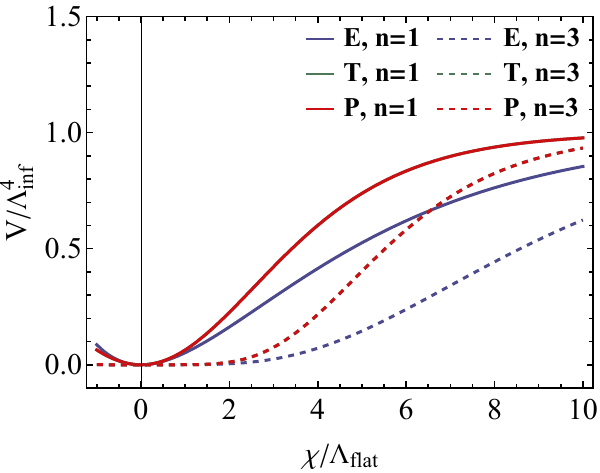}
        \caption{$\alpha = 10$}
    \end{subfigure}
    \caption{The $\alpha$-attractor potentials as functions of $\frac{\chi}{\Lambda_\fl}$, for different values of $\alpha$ and $n$ in the E-, T-, P-models defined in \cref{eq:alpha-attractors}.
    }
    \label{fig:alphaPot}
\end{figure}
Substituting \cref{eq:Kahler,eq:ChiAndTheta} in \cref{eq:FlavonPotentialK} yields the canonically normalized Lagrange density
\begin{equation}
\label{eq:LchiTheta}
    \mathcal{L}_{\sigma, \theta} ~=~ \frac{\partial_\mu \chi \partial^\mu \chi }{2} +  f(v_\phi)\,\frac{\partial_\mu \theta \partial^\mu \theta }{2} - V\left( \chi,\alpha, n,\Lambda_\mathrm{inf},\Lambda_\fl\right)+\mathcal{L_{\rm int}}\; ,
\end{equation}
for the three different potentials in \cref{eq:alpha-attractors}, where  $f(v_\phi)$ is some function that depends on $v_\phi$ and $\mathcal{L_{\rm int}}$ are the interactions between $\chi$ and $\partial\theta$.  We can expand the flavon field as in \cref{eq:flavonVev} and expand \cref{eq:ChiAndTheta} for small $\theta$ and $\chi$. In all three cases, we get that, to leading order, the relation between $\sigma$ and $\chi$ is given by
\begin{equation}
\label{eq:SigmaAndChi}
    \sigma ~\simeq~ \frac{v_\phi}{2\Lambda_\fl}\sqrt{\frac{2}{3\alpha}} \chi\;.
\end{equation}
Hence, to leading order, $\chi$ is the canonically-normalized field of the flavon field $\sigma$. In other words, the field $\chi$ serves the role of an inflaton field and a flavon field, or an \textit{inflavon}. We can obtain the couplings of $\chi$ to the \ac{SM} and the right-handed neutrinos. First, substituting \cref{eq:flavonVev} in \cref{eq:BeforeFlavorL} we obtain
\begin{equation}
\label{eq:sigma}
    -\mathcal{L}_{\sigma , \, \mathrm{int}} ~\supset~ \frac{g_{ij}^{u}}{v_\phi} \sigma \bar Q_i \widetilde H u_j + \frac{g_{ij}^{d}}{v_\phi} \sigma \bar Q_i H d_j + \frac{g_{ij}^{e}}{v_\phi} \sigma \bar L_i  H E_j + \frac{g_{ij}^{D}}{v_\phi} \sigma  \bar L_i \widetilde H N_j + \mathrm{h.c.} \; ,
\end{equation}
where 
\begin{equation}
    \label{eq:gij}
    g_{ij}^{a} = |n_{ij}^{a}|\lambda^{|n_{ij}^{a}|}\alpha_{ij}^{a}\; . 
\end{equation}
Then, substituting \cref{eq:SigmaAndChi} in \cref{eq:sigma}, we obtain the couplings between the inflavon $\chi$ and the \ac{SM} 
\begin{align}
\label{eq:chiSM}
    -\mathcal{L}_{\chi, \mathrm{int}}  ~\supset~  &\left( \frac{g_{ij}^{u}}{2\Lambda_\fl}\sqrt{\frac{2}{3\alpha}} \chi \right) \bar Q_i \tilde H u_j + \left( \frac{g_{ij}^{d} }{2\Lambda_\fl}\sqrt{\frac{2}{3\alpha}} \chi \right) \bar Q_i H d_j \nonumber \\[1em]
    &  + \left( \frac{g_{ij}^{e} }{2\Lambda_\fl}\sqrt{\frac{2}{3\alpha}} \chi \right) \bar E_i \tilde H L_j+ \left( \frac{ g_{ij}^{D} }{2\Lambda_\fl}\sqrt{\frac{2}{3\alpha}} \chi \right)  L_i H N_j + \mathrm{h.c.}\; .
\end{align}
We comment that the mixing term between the flavon and the Higgs $|H|^2 |\phi|^2$ is unavoidable. However, for the sake of simplicity, we assume that this coupling vanishes. If the magnitude of this mixing term were sufficiently large, it would lead to the flavon predominantly decaying into a pair of Higgs bosons after \ac{EW} symmetry breaking. In \Cref{app:Higgs-Flavon-Mixing}, we estimate the upper bound on this coupling required for the validity of our analysis. Additionally, we also argue that one-loop corrections do not induce a dominant decay channel into a Higgs boson pair.

\subsection*{Inflavon as dark matter and isocurvature constraints}

By expanding \cref{eq:ChiAndTheta} around $\phi = v_\phi$, we see that $\rho$ is given by
\begin{equation}
    \rho  \simeq  \frac{v_\phi}{2 \Lambda_{\fl}} \sqrt{\frac{2}{3 \alpha}}\theta\;,
\end{equation}
for all three inflationary models. Here the field $\rho$ is the so-called \textit{flaxion}~\cite{Ema:2016ops,Calibbi:2016hwq}. We can identify the axion decay constant with 
\begin{equation}
\label{eq:ftheta}
    f_{\rho } = \sqrt{3\alpha} \frac{\Lambda_{\fl}}{ N_{\mathrm{DW}}}\;,
\end{equation}
where $N_{\mathrm{DW}}:=\mathrm{Tr}\left(n^{u}+n^{d}\right)$ is the domain wall number, and $n^{u}$ and $n^{d}$ are the \ac{FN} charges defined in \cref{eq:nij}. $N_{\mathrm{DW}}$ is of order $\mathcal{O}\left(10\right)$ for order one \ac{FN} charges. 

The axion decay constant is constrained by flavor-changing processes such as $K^{+}\to\pi^{+}+\rho$ or $\mu\to e+\rho+\gamma$. These bounds have been computed in~\cite{Ema:2016ops,Calibbi:2016hwq} for the case where the imaginary part of the flavon is identified as the \ac{QCD} axion. In our case, the axion decay constant is of order of $\mathcal{O}\left(10^{-3}-1\right) M_{\pl}$ for the values of $\alpha$ considered in this work. Hence, it can be easily checked that these bounds are satisfied. 

Finally, the present energy density of the \ac{QCD} axion $\Omega_\rho h^2$ is given by~\cite{Ema:2016ops}
\begin{equation}
    \Omega_\rho h^2 = 0.18\,\theta_i^2\,\left(\frac{f_\rho}{10^{12}\,\mathrm{GeV}}\right)^{1.19}\;,
\end{equation}
where $\theta_i$ is the initial axion misalignment angle. If the \ac{QCD} axion accounts for the observed dark matter abundance, then for an axion decay constant of order $\mathcal{O}\left(10^{-3}-1\right) M_{\pl}$, the initial misalignment angle must lie within the range $\mathcal{O}\left(10^{-4}-10^{-2}\right)$. Nevertheless, note that the axion oscillations can produce isocurvature fluctuations. These are highly constrained by Planck data~\cite{Planck:2015sxf}. These have been studied in the context of a flaxion~\cite{Ema:2016ops}. For a canonically-normalized axion kinetic term, the Hubble rate during inflation $H_{\mathrm{inf}}$ must satisfy the constraint given by
\begin{equation}
\label{eq:isocurvatureConstraints}
    H_{\mathrm{inf}} \lesssim  3 \times 10^{7}\,\mathrm{GeV}\,\theta_i^{-1}\left( \frac{10^{12}\,\mathrm{GeV}}{f_\rho}\right)^{0.19}\;.
\end{equation}
For the $E-$model, the axion field $\theta$ in \cref{eq:ChiAndTheta} is canonically normalized. Accordingly, we check if \cref{eq:isocurvatureConstraints} is satisfied. For the models considered in this work, the bound is not satisfied if the flaxion constitutes all the \ac{DM} abundance. However, certain scenarios can avoid these constraints; see e.g.\ \cite{Berbig:2024ufe}. A detailed analysis of the flaxion abundance and its isocurvature constraints, in this context, is deferred to future work.

\section{\ac{CMB} imprints from $\alpha-$attractor inflaton potentials}
\label{sec:CMB-alpha}

The inflationary reheating era has been studied in the past, in general,  \cite{Kofman:1994rk,Kofman:1997yn}, and, particularly, also in the context of $\alpha$-attractor models of inflation \cite{Kallosh:2013lkr,Kallosh:2013hoa,Kallosh:2013yoa,Kallosh:2013pby,Kallosh:2013maa,Galante:2014ifa}. We revisit the relations between the parameters in the $\alpha-$attractor potentials from \cref{eq:alpha-attractors} and \ac{CMB} observables: the amplitude of the scalar perturbations $A_s$, the tensor-to-scalar ratio $r$ and the spectral index $n_s$. The explicit relations between the different $\alpha-$attractor potentials of \cref{eq:alpha-attractors} and the aforementioned observables are presented in \Cref{app:AlphaAtractors}. In this Section, we just outline the general method that we use in this work. The \ac{CMB} observables are evaluated at some reference scale, typically the pivot scale $k=0.05 \,$Mpc$^{-1}$. 
A general inflaton potential $V(\chi)$ can be characterized through the slow-roll parameters $\epsilon$ and $\eta$ defined by~\cite{Lyth:2009imm}:
\begin{align}
\label{eq:SlowRoll}
    \epsilon\left( \chi,\alpha, n,\Lambda_\mathrm{inf},\Lambda_\fl \right) &~:=~ \frac{1}{2}M_{\pl}^2\left( \frac{\partial_\chi V(\chi,\alpha, n,\Lambda_\mathrm{inf},\Lambda_\fl)}{V\left( \chi,\alpha, n,\Lambda_\mathrm{inf},\Lambda_\fl \right) }\right)^2 \,, \nonumber \\[1em]
    \eta \left( \chi,\alpha, n,\Lambda_\mathrm{inf},\Lambda_\fl \right) &~:=~ M_{\pl}^2\left(\frac{\partial_\chi^2 V\left(\chi,\alpha, n,\Lambda_\mathrm{inf},\Lambda_\fl\right)}{V(\chi,\alpha, n,\Lambda_\mathrm{inf},\Lambda_\fl)}\right) \,,
\end{align}
where $M_{\pl} = 2.4 \times 10^{18}\,$GeV is the reduced Planck mass. The \ac{CMB} observables $n_s$, $r$ and $A_s$ are then defined as
\begin{align}
    n_s \left( \chi_k,\alpha, n,\Lambda_\mathrm{inf},\Lambda_\fl \right) &~:=~1-6\epsilon \left( \chi_k,\alpha, n,\Lambda_\mathrm{inf},\Lambda_\fl  \right) +2\eta \left( \chi_k,\alpha, n,\Lambda_\mathrm{inf},\Lambda_\fl \right)\;,\label{eq:ns}\\[1em]
    r\left( \chi_k,\alpha, n,\Lambda_\mathrm{inf},\Lambda_\fl \right) &~:=~16 \epsilon \left( \chi_k,\alpha, n,\Lambda_\mathrm{inf},\Lambda_\fl \right)\;,\label{eq:r}\\[1em]
    A_s \left( \chi_k,\alpha, n,\Lambda_\mathrm{inf},\Lambda_\fl \right) &~:=~\frac{H^2\left( \chi_k,\alpha, n,\Lambda_\mathrm{inf},\Lambda_\fl \right)}{M_{\pl}^2}\frac{2}{\pi^2}\frac{1}{r }\;, \label{eq:As}
\end{align}
where the $\chi_k := \chi\left(t_k\right)$ is the field value $\chi$ at the time $t_k$ when the Fourier mode $k$ crosses the horizon. This corresponds to the time when the comoving wavenumber $k$ becomes comparable to the Hubble rate, satisfying the condition $k = R H$ where $R$ is the scale factor. Hence, the value of the Fourier mode $\chi_{\vec k}$ of the field $\chi$ remains constant after $t_k$. Furthermore, we can also define the derivatives of the spectral index $n_s$ with respect to the scale $k$. The first one of these is the so-called \textit{running of the scalar  spectral index} $\alpha_s$. In \cref{eq:alphasdlnk,eq:alphasetaepsilonxi,eq:alphsSexplicit} in \Cref{app:AlphaAtractors}, we present explicit expressions for $\alpha_s$. However, in the numerical analysis throughout this work, we find that $\alpha_s$ is too small to be constrained by current experimental data~\cite{Planck:2018vyg,Choudhury:2025vso}. Consequently, we choose to neglect $\alpha_s$ and higher-order derivatives throughout our work. Using the slow-roll approximation $\epsilon\ll 1$ during inflation yields
\begin{equation}
   \label{eq:Vchik}
    H^2\left(\chi_k,\alpha, n,\Lambda_\mathrm{inf},\Lambda_\fl \right) ~\simeq~ \frac{V\left(\chi_k,\alpha, n,\Lambda_\mathrm{inf},\Lambda_\fl \right)}{3 M_{\pl}^2}\;.
\end{equation}
Thus, substituting \cref{eq:Vchik} in \cref{eq:As} gives
\begin{equation}
\label{eq:As2}
    A_s \left( \chi_k,\alpha, n,\Lambda_\mathrm{inf},\Lambda_\fl \right) ~:=~\frac{2V\left( \chi_k,\alpha, n,\Lambda_\mathrm{inf},\Lambda_\fl \right)}{3 \pi^2M_{\pl}^4}\frac{1}{r \left( \chi_k,\alpha, n,\Lambda_\mathrm{inf},\Lambda_\fl \right) }\;.
\end{equation}
We can solve for $\chi_k = \chi_k\left(n_s,n, \alpha,\Lambda_\mathrm{inf},\Lambda_\fl\right)$ from \cref{eq:ns}. 
Furthermore, using this relation for $\chi_k$ in the second line of \cref{eq:r} together with the observed value of the amplitude of the scalar perturbations $A^{\mathrm{obs}}=2.090\times 10^{-9}$~\cite{Planck:2018vyg} in \cref{eq:As2} gives the relations 
\begin{align}
\label{eq:LastRAndLambdaInfRelations}
     r~&=~r\left(n_s,\alpha, n,\Lambda_\fl \right)\;,\nonumber\\[1em]
     \Lambda_\mathrm{inf} ~&=~ \Lambda_\mathrm{inf}\left(n_s, \alpha, n, \Lambda_\fl\right)\;.
\end{align}
Since $\Lambda_\mathrm{inf} = \kappa v_\phi^{4n} = \kappa (0.17)^{4n} \Lambda_\FN^{4n}$ (see \cref{eq:FlavonPotentialK}), then, this imposes a relation for $\Lambda_\FN$ and $\kappa$ too. \Cref{eq:LastRAndLambdaInfRelations} indicates that all the quantities derived from the $\alpha-$attractor potentials in \cref{eq:alpha-attractors} can, at most, depend solely on $n_s,\alpha, n,\Lambda_\fl$. We note that three of these parameters are parameters of the $\alpha-$attractor potential in \cref{eq:alpha-attractors}: $\alpha$, $n$, and $\Lambda_\fl$. On the other hand, $n_s$ is an observable. We write the dependence with respect to these four variables to show that after completely fixing the inflaton potential, then all the quantities in this Section depend only on one observable $n_s$. Thus, from now on, we will ignore the dependence on these variables.

\begin{table}[t!]
	\centering
	\begin{subtable}[t]{0.38\textwidth}
		\centering
		\begin{tabular}{ll}
			\toprule
			observables          & best-fit values  \\
			\midrule
			$m_\mathrm{e}/m_\mu$ & $0.00474\pm0.00004$ \\
			$m_\mu/m_\tau$       & $0.0588 \pm 0.0005$ \\
            $m_u/m_c$            & $0.00193\pm0.00060$ \\
            $m_c/m_t$            & $0.00280\pm 0.00012$ \\
            $m_d/m_s$            & $0.0505\pm0.0062$   \\
            $m_t/\text{GeV}$     & $89.213 \pm 2.219$ \\
            $m_b/\text{GeV}$     & $0.965 \pm 0.011$ \\
            $\sin^2\theta_{12}^q$& $0.0511\pm0.0003$ \\
            $\sin^2\theta_{13}^q$& $1.22\pm 0.09 \times 10^{-5}$ \\
            $\sin^2\theta_{23}^q$& $0.00165\pm0.00005$\\
			\bottomrule
		\end{tabular}
		\caption{Quark sector.}
		\label{tab:ExpDataLeptonMasses}
	\end{subtable}%
	\hfill
	\begin{subtable}[t]{0.61\textwidth}
		\centering
		\begin{tabular}{lll}
			\toprule
			observables & Normal Ordering & Inverted Ordering \\
			\midrule
            $\Delta m_{21}^2/10^{-5}\text{eV}^2$ &
            $7.41^{+0.21}_{-0.20}$ &
            $7.41^{+0.21}_{-0.20}$
            \\
			$\Delta m_{21}^2 / \Delta m_{31}^2$ & $0.0296^{+0.0012}_{-0.0011}$ &
            $-0.0298^{+0.0011}_{-0.0011}$
            \\[4pt]
			$\sin^2\theta_{12}$                 & 
            $0.307^{+0.012}_{-0.011}$ &
            $0.307^{+0.012}_{-0.011}$ \\[4pt]
			$\sin^2\theta_{13}$                 & 
            $0.02224^{+0.00056}_{-0.00057}$ &
            $0.02222^{+0.00069}_{-0.00057}$\\[4pt]
			$\sin^2\theta_{23}$                 & 
            $0.454^{+0.019}_{-0.016}$ &
            $0.568^{+0.016}_{-0.021}$\\[4pt]
			$\delta_{\CP}^{\ell}/^\circ$           & 
            $232^{+39}_{-25}$ &
            $273^{+24}_{-26}$\\[2pt]
			\bottomrule
		\end{tabular}
		\caption{Lepton sector.}
		\label{tab:ExpDataNeutrinoMixing}
	\end{subtable}
	\caption{
		Experimental best-fit values for the masses and mixing parameters of the \ac{SM}. The left panel shows observables for quark and charged lepton sector evaluated at the GUT scale~\cite{Ross:2007az}, while the right shows neutrino mixing observables from NuFIT v5.3, taking the Super-Kamiokande data into account~\cite{Esteban:2020cvm}.
	}
	\label{tab:ExpData}
\end{table}

We can further obtain a relation between the \ac{CMB} observables and a given particle physics model as shown in~\cite{Drewes:2015coa}. First, recall that the end of the reheating period is the instant at which the Hubble rate matches the decay width of the inflation, that is 
\begin{equation}
\label{eq:Gamma=H}
    \Gamma_{\chi}(t_\mathrm{RH} )~\approx~ H_\mathrm{RH}\; ,
\end{equation}
where $\Gamma_\chi$ is the total decay width\footnote{Note that from \cref{eq:chiSM} we see that the inflaton only interacts through $5-$dimensional terms in the Lagrangian. This could give either $3-$body decays or $2$-to-$2$ scattering processes. The scattering interaction rate goes as $\sim \frac{T^3}{\Lambda_\fl^2}$ while the decay width goes as $\sim \frac{m_\chi^3}{\Lambda_\fl^2}$. Thus, for $m_\chi \gg T$, we can safely ignore the scattering process contribution. We will see in \Cref{sec:FNFaceCosmology} that this is the case for this work.} of the inflavon $\chi$ and $H_\mathrm{RH}$ is the Hubble rate at reheating \footnote{Here we adopted the condition in \cref{eq:Gamma=H} following~\cite{Drewes:2015coa}. However, often other conditions such as $\Gamma_{\chi} (t_{\re}) \approx 3  H_{\re}$ might be used. Either choice is expected to yield similar results.}. On one hand, the left-hand side of \cref{eq:Gamma=H} depends solely on the particle physics model. On the other hand, the right-hand side of \cref{eq:Gamma=H} depends only on $n_s,\alpha, n,\Lambda_\fl$. The Hubble rate during reheating is given by
\begin{equation}
\label{eq:Hre}
    H_\mathrm{RH}  ~=~ \frac{1}{M_{\pl}}\left(\frac{\rho_{\mathrm{end}}}{3} \right)^{1/2} \ee^{-3 \left( 1 + \overline{w}_{\mathrm{RH}}\right) N_{\mathrm{RH}}/2}
\end{equation}
where $\rho_\mathrm{end}$ is the energy density of the inflavon at the end of inflation. It is given by
\begin{equation}\label{eq:rhoEndGeneral}
    \rho_{\text{end}}  ~\simeq~ \left( 1 + \frac{\epsilon\left(\chi_\mathrm{end} \right)}{3} \right) V \left( \chi_\mathrm{end} \right) = \frac{4}{3 }V \left( \chi_\mathrm{end} \right) \; ,
\end{equation}
where $\chi_\mathrm{end}$ is the inflavon value at the end of inflation when $\epsilon\left(\chi_\mathrm{end}\right) = 1$.\footnote{Note that the general condition for inflation to end is $\mathrm{min}(t_1, t_2)$ such that $\epsilon (t_1) = 1$ and $\left| \eta (t_2 ) \right| =1$.} We have verified that, in all cases considered in this work, we get that the condition $\epsilon = 1$ is satisfied before $\left| \eta \right| = 1$. Since all the $\alpha-$attractor potentials behave like $V \propto \chi^{2n}$ during reheating, then under certain assumptions (see~\cite{Shtanov:1994ce,Cembranos:2015oya} for details), the average equation of state parameter $\overline{w}_\mathrm{RH}$ during reheating is given by\footnote{Here, the contributions of the inflaton fluctuations, or in general, density fluctuations, to $w_{\rm re}$ have been neglected. However, these can give corrections to the equation of state in \cref{eq:omega_re}, see~\cite{Cembranos:2015oya}. We do not expect these corrections to change significantly our results as we study reheating around the minimum.}
\begin{equation}
\label{eq:omega_re}
    \overline{w}_{\mathrm{RH}} ~\approx~ \frac{n - 1}{n +1} \, .
\end{equation}
 The number of e-folds $N_{\mathrm{RH}}$ during reheating is given by
\begin{align}
    \label{eq:Nre}
    N_{\mathrm{RH}}  ~&=~ \frac{4}{3 \overline{w}_{\mathrm{RH}}-1}\bigg[N_k +\ln\left(\frac{k}{R_0 T_0}\right) + \frac{1}{4}\ln \left( \frac{40}{\pi^2 g_{*}}\right) + \frac{1}{3}\ln\left( \frac{11g_{*}}{43} \right) \nonumber\\
    &- \frac{1}{2}\ln \left( \frac{\pi^2 M_{\pl}^2 r A_s^{\mathrm{obs}}}{2 V^{1/2}\left(\chi_\mathrm{end}\right)} \right)\bigg]\; ,
\end{align}
where $g_{*}$ is the relativistic degrees of freedom, $T_0$ is the temperature now, $R_0$ the scale factor now and $N_{k}$ is the number of e-folds between the moment a perturbation with wavenumber $k$ crosses the horizon and the end of inflation. We fix $g_{*} = 109.375$ to account for the $\ac{SM}$ and three right-handed neutrinos. In addition, we fix the scalar $R_0 = 1 $ and the temperature today as $T_0 = 2.7$ K. The number of e-folds $N_k$ is given by 
\begin{align}
   \label{eq:Nk}
    N_k  & ~=~ \log \left( \frac{R_{\mathrm{end}}}{R_k}\right) \simeq -\frac{1}{M_{pl}^2}\int_{\chi_k}^{\chi_{\mathrm{end}}}d\chi\; \frac{V \left(\chi\right) }{\partial_\chi \left(\chi\right) } \; .
\end{align}
Thus, using \cref{eq:Hre,eq:rhoEndGeneral,eq:omega_re,eq:Nre,eq:Nk} in \cref{eq:Gamma=H} yields~\cite{Drewes:2017fmn}
\begin{equation}
\label{eq:GammaKey}
    \Gamma_\chi (t_\mathrm{RH}) ~=~ \frac{1}{M_{\pl}}\left(\frac{4 V\left( \chi_\mathrm{end} \right) }{9}\right)^{1/2}\ee^{-3 \left(1 +\overline{w}_{\mathrm{RH}}\right) N_{\mathrm{RH}} /2}\, ,
\end{equation}
where we remark that the right-hand side of \cref{eq:GammaKey} depends only on $n_s,\alpha, n,\Lambda_\fl$. In our case, the left-hand side depends on the dimensionless couplings $\alpha_{ij}^{a}$ (see \cref{eq:BeforeFlavorL}) of the different sectors of the \ac{SM}, the $\U{1}_\FN$ charges and $v_\phi$. Since $v_\phi$ is related with $\Lambda_\mathrm{inf}$ through \cref{eq:LastRAndLambdaInfRelations}, then the left-hand side of \cref{eq:GammaKey} depends only on $n_s,\alpha, n,\Lambda_\fl$ and $\kappa$. Note that this is different from~\cite{Drewes:2017fmn} where the decay width of the inflaton is assumed to not depend explicitly on $n_s,\alpha, n$ nor $\Lambda_\fl$. Thus, by fixing the inflaton potential completely by specifying $\alpha$, $n$ and $\Lambda_{\fl}$, \cref{eq:GammaKey} can be solved to obtain $n_s$. Then, we can calculate $r$ and $\Lambda_{\inf}$ from \cref{eq:LastRAndLambdaInfRelations}.

In addition, we can obtain the reheating temperature. We assume that the conversion of the energy density of $\chi$ into the energy density of the thermal bath occurs instantaneously at the end of reheating yielding
\begin{equation}\label{eq:AssumptionInstantaneous}
    \rho_\mathrm{RH} ~=~ \frac{\pi^2}{30}g_{*}T_\mathrm{RH}^4\;,
\end{equation}
where $\rho_\mathrm{RH}$ is the energy density of $\chi$ at the end of reheating and $T_\mathrm{RH}$ is the reheating temperature. Using \cref{eq:AssumptionInstantaneous} with \cref{eq:GammaKey} gives the reheating temperature
\begin{equation}
    \label{eq:Tre}
    T_{\mathrm{RH}} ~=~ \ee^{-3 \frac{1+ \overline{w}_{\mathrm{RH}}}{4} N_{\mathrm{RH}}}\left( \frac{40 V \left(\chi_\mathrm{end} \right)}{g_{*} \pi^2}\right)^{1/4} \; .
\end{equation}

\medskip

\begin{table}[t]
\centering
\begin{subtable}[h]{\textwidth}
\centering
\begin{tabular}{c|ccccccccccc}
Model & $Q_1$ & $Q_2$ & $Q_3$ & $U_1$ & $U_2$ & $U_3$ & $D_1$ & $D_2$ & $D_3$ & Up Quark Hierarchy & Down Quark Hierarchy \\ \hline
1  & 0 & 1 & 0 & 5 & 1 & 3 & 3 & 0 & 3 & $(1,\lambda^3,\lambda^5)$ & $(1,\lambda^2,\lambda^3)$\\
2   & 1 & 0 & 1 & 5 & 1 & 3 & 3 & 1 & 3  & $(1,\lambda^2,\lambda^5)$ & $(1,\lambda^2,\lambda^3)$\\
\end{tabular}
\end{subtable}

\begin{subtable}[h]{\textwidth}
\centering
\begin{tabular}{c|ccccccccc}
Model & Order & $L_1$ & $L_2$ & $L_3$ & $E_1$ & $E_2$ & $E_3$ & Charged Lepton Hierarchy & Neutrino Hierarchy \\ \hline
$\circ$ & NO & 2 & 3 & -3 & 0 & 0 & -3 & $(1,\lambda^2,\lambda^3)$ & $(1,\lambda,\lambda^6)$\\
$\circ'$ & NO & 1 & -3 & 2 & 0 & 0 & 1 & $(1,\lambda^2,\lambda^3)$& $(1,\lambda,\lambda^5)$ \\
$\hat{\circ}$ & IO & 0 & 2 & -1 & 3 & -1 & 0 & $(1,\lambda^2,\lambda^3)$ & $(1,1,\lambda^2)$\\
$\tilde{\circ}$ & IO & -3 & -1 & -2 & 0 & 1 & 3 & $(1,\lambda^3,\lambda^6)$& $(1,1,\lambda^7)$ \\
\end{tabular}
\end{subtable}

\caption{$\U{1}_\FN$ charges for each model, where the charge for $E$, the charged lepton singlet, are all chosen to have charge -3. The flavon is not coupled to the right-handed neutrino ($N$), i.e., right-handed neutrino are not charged under $\U{1}_\FN$. }
\label{tab:models}
\end{table}

Before proceeding we will make certain simplified assumptions: we have assumed perturbative reheating. There can be that, due to the inflavon couplings to the SM particles, this simple picture breaks down and non-perturbative particle production may become important. Some discussion of non-perturbative effects can be found in ~\cite{Drewes:2017fmn,Drewes:2019rxn}\footnote{These studies show that, for renormalizable couplings, non-perturbative effects can be ignored if these couplings are smaller than $10^{-6}$. In our case, we consider non-renormalizable couplings. Hence, we expect these bounds to be satisfied due to the suppression of non-renormalizable couplings.}. We assume that the Coleman-Weinberg loop corrections to the inflavon potential (during reheating) generated by the inflavon couplings do not destabilize it and also do not significantly modify shape of the potential, so as to impact the inflationary CMB observables. We ignore the effects of inflavon decay $\Gamma$ during inflation, their contributions to CMB observables, for instance on the primordial non-gaussianities. In addition, we  assume the gravitational production of SM radiation bath during inflation to be negligible, see Ref.\cite{Ghoshal:2024gai} for a discussion when it is justified. Finally, we assume instant and local thermal equilibrium to have been reached during the reheating period.

\section{Froggatt-Nielsen faces cosmology constraints}
\label{sec:FNFaceCosmology}
The decay width of the inflaton $\chi$ into three particles is obtained from the Lagrange density in \cref{eq:chiSM}. Using that  the inflaton mass is much bigger than the mass of the rest of the particles, the decay width is given by
\begin{equation}
\label{eq:DecayWidthThreeBody}
    \Gamma_{ij}^{a} ~ = ~\frac{N_c}{3} \frac{\left|g_{ij}^{a}\right|^2}{64 \pi^3}\left|\frac{1}{2 \Lambda_\fl} \sqrt{\frac{2}{3\alpha}} \right|^2 m_\chi^3\;,
\end{equation}
where $N_c$ is the number of colors, $m_\chi$ is the mass of the inflaton and we fix $\Lambda_{\fl} = M_{\pl}$ for the rest of this work. The mass is obtained from the $\alpha-$attractor potential in \cref{eq:alpha-attractors} and is given by
\begin{equation}
\label{eq:mchi2}
    m_\chi^2 ~=~ \begin{cases}
        \frac{4 \Lambda_{\text{inf}}^4}{3 \alpha \Lambda_\fl^2} \, \text{ for } n=1\, ,\\
        0, \,\text{ otherwise}\, .
    \end{cases}
\end{equation}
Naively, using \cref{eq:mchi2} in \cref{eq:DecayWidthThreeBody} yields a vanishing decay width for the inflaton for $n\neq 1$. However, it has been shown~\cite{Ichikawa:2008ne,Garcia:2020wiy,Barman:2023ktz} that it is still possible to calculate the energy transfer from the inflaton to the rest of the other particles. Hence, we distinguish two different cases: $n=1$ and $n \neq 1$.\footnote{Note that for $n=2$, the inflavon behaves like radiation. Then, we have $N_{\re} = 0$ and the \ac{CMB} observables can be obtained without calculating the energy transfer of the inflavon to the rest of the sectors in the Universe.}

Furthermore, one may ask whether the inflavon could decay into the axion field $\theta$, since a coupling of the form $g(\chi)\chi\left(\partial\theta\right)^2$ can arise in \cref{eq:LchiTheta}. For instance, for the $T-$model, one finds $g\left(\chi\right)\sim 1 / \Lambda_{\fl}$. Thus, leading to a decay width of the form
\begin{equation}
    \Gamma_{\chi \to \theta \theta} \sim \frac{m_{\chi}^3}{\Lambda_{\fl}^2}\;,
\end{equation} 
which could become comparable to the contribution discussed in \cref{eq:DecayWidthThreeBody}. To avoid this channel, the $\U{1}_{\FN}$ can be gauged. In this case, then the mass of the $\U{1}_{\FN}$ gauge boson $A_{\mu}^{'}$ is given by $m_{A'} = g v_\phi$, where $g$ is the gauge coupling constant. Since the inflavon mass scales  as $\sim \frac{\Lambda_{\inf}^2}{\Lambda_{\fl}}$ and the expectation value is given by $v_{\phi} \sim 0.1 \Lambda_{\inf}$, it follows that, for $\Lambda_{\fl} = M_{\pl} \gg  \Lambda_{\inf}$, the gauge boson is heavier than the inflavon. Consequently, the decay channel $\chi \to A_{\mu}^{'} A_{\mu}^{'}$ is kinemaically forbidden.  In that case, one must ensure that the anomalies of the $\U{1}_{\FN}$ symmetry cancel. Here, we focus on illustrating how different charge assignments affect the \ac{CMB} signatures, while leaving a detailed analysis of anomaly cancellation requirements to future work.

We analyze a large number of different \ac{FN} charges assignments. For each set of charges, we perform a scan to identify the best-fit configuration that matches the experimental flavor parameters (see \Cref{tab:ExpData}). Additionally, we verify that the models are not fine-tuned. Specifically, we require the coupling constants $\alpha_{ij}$ in \cref{eq:BeforeFlavorL} to be of the same order within each sector, and that mass hierarchies then arise solely from small expansions in $\lambda$. However, we explicitly analyze the predicted inflationary parameters $n_s$ and $r$ only for a few representative models, summarized in \Cref{tab:models}. We present two quark models, labeled by $1$ and $2$; two \ac{IO} lepton models, labeled by a $\hat{\circ}$ and $\tilde{\circ}$ respectively; and two \ac{NO} lepton models, labeled $\circ$ and $\circ'$.

\subsection{$\alpha-$attractor potentials for $n=1$}
\label{subsec:alphaNEquals1}

\begin{figure*}[t!]
    \centering
    \begin{subfigure}[t]{\textwidth}
        \centering
        \includegraphics{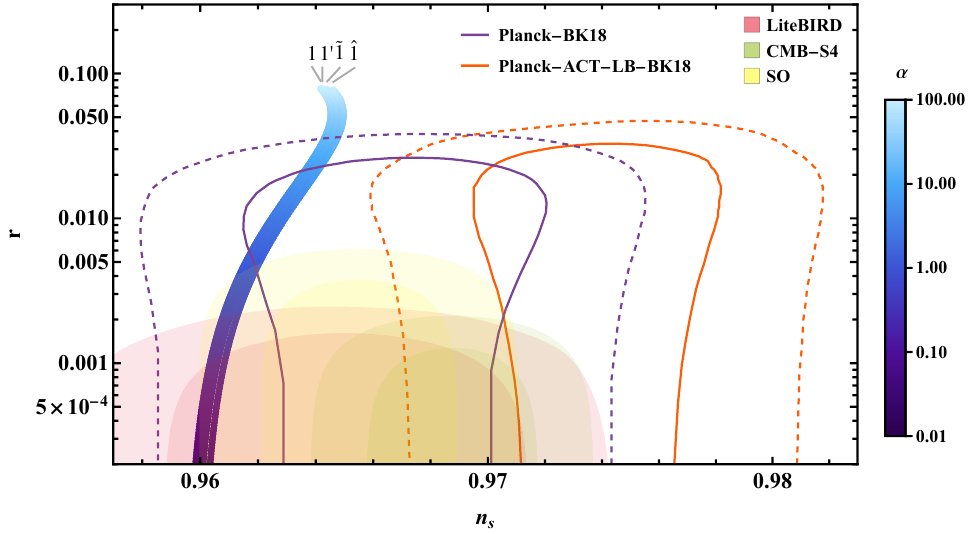}
        \caption{Tensor-to-scalar ratio $r$ vs. primordial tilt $n_s$ plot and for E-model in \Cref{tab:models} for quark model 1}
    \end{subfigure}%
    ~ \\
    \begin{subfigure}[t]{\textwidth}
        \centering
        \includegraphics{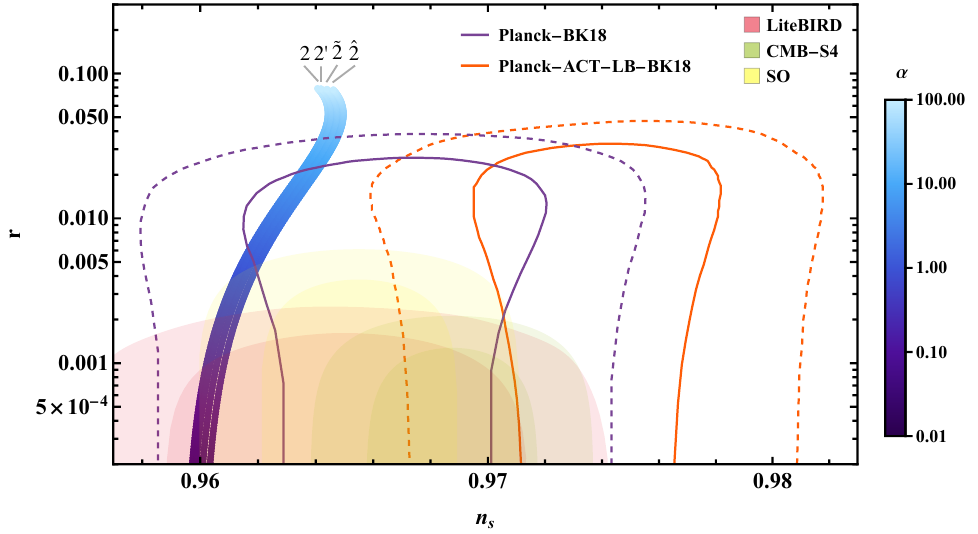}
        \caption{Tensor-to-scalar ratio $r$ vs. primordial tilt $n_s$ plot and for E-model in \Cref{tab:models} for quark model 2}
    \end{subfigure}
    \caption{\label{fig:nsrPlotEmodel}Results for E-Model. The solid and dashed lines are the $1-$sigma and $2-$ sigma current constraints from \textit{Planck}, \textit{BK15}, and \textit{ACT} ~\cite{Planck:2018vyg,BICEP2:2018kqh}. The light and dark shaded regions represent future experimental constraints from \textit{LiteBIRD}, \textit{CMB-S4}, and \textit{SO}\cite{LiteBIRD:2022cnt,CMB-S4:2016ple,SimonsObservatory:2018koc}.}
\end{figure*}
\begin{figure*}[t!]\ContinuedFloat
    \centering
    \begin{subfigure}[t]{\textwidth}
        \centering
        \includegraphics{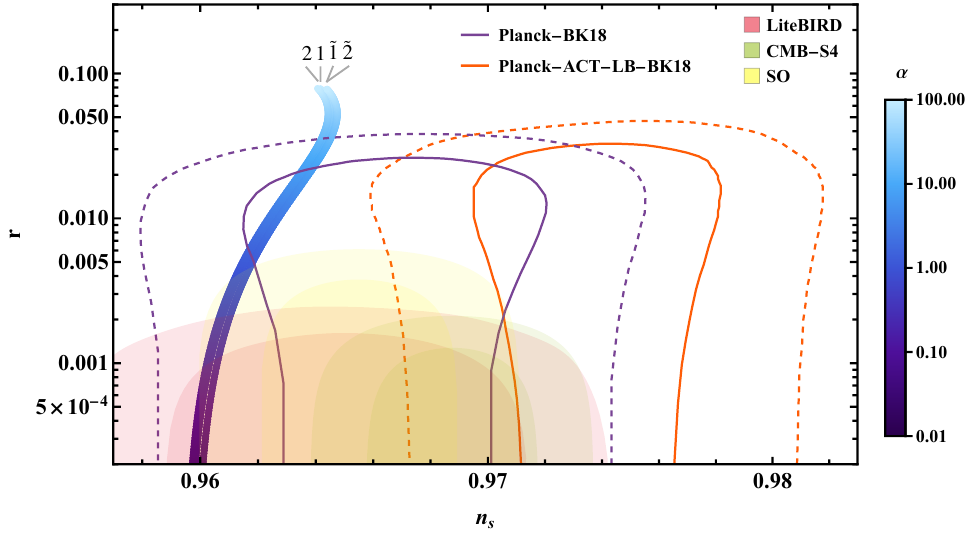}
        \caption{Tensor-to-scalar ratio $r$ vs. primordial tilt $n_s$ plot and for E-model in \Cref{tab:models} for model 1, 2(NO) and $\tilde{1}$, $\tilde{2}$(IO)}
    \end{subfigure}
    ~ \\
    \begin{subfigure}[t]{\textwidth}
        \centering
        \includegraphics{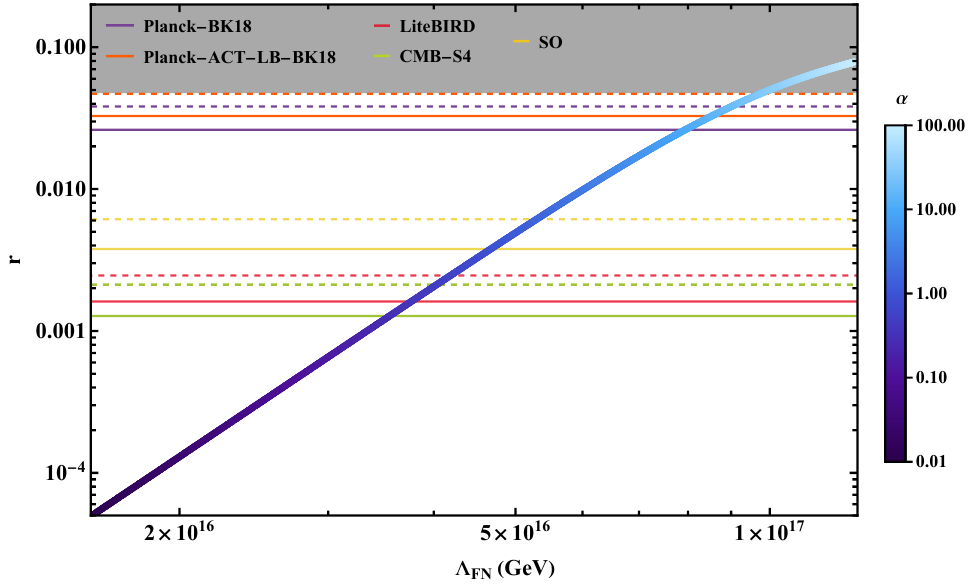}
        \caption{$r$ vs. \ac{FN} scale $\Lambda_{\FN}$ plot for E-model.}
    \end{subfigure}
    \caption{Results for E-Model. The solid and dashed lines are the $1-$sigma and $2-$ sigma current constraints from \textit{Planck}, \textit{BK15}, and \textit{ACT} ~\cite{Planck:2018vyg,BICEP2:2018kqh}. The light and dark shaded regions represent future experimental constraints from \textit{LiteBIRD}, \textit{CMB-S4}, and \textit{SO}\cite{LiteBIRD:2022cnt,CMB-S4:2016ple,SimonsObservatory:2018koc}. }
\end{figure*}

\begin{figure*}[t!]
    \centering
    \begin{subfigure}[t]{\textwidth}
        \centering
        \includegraphics{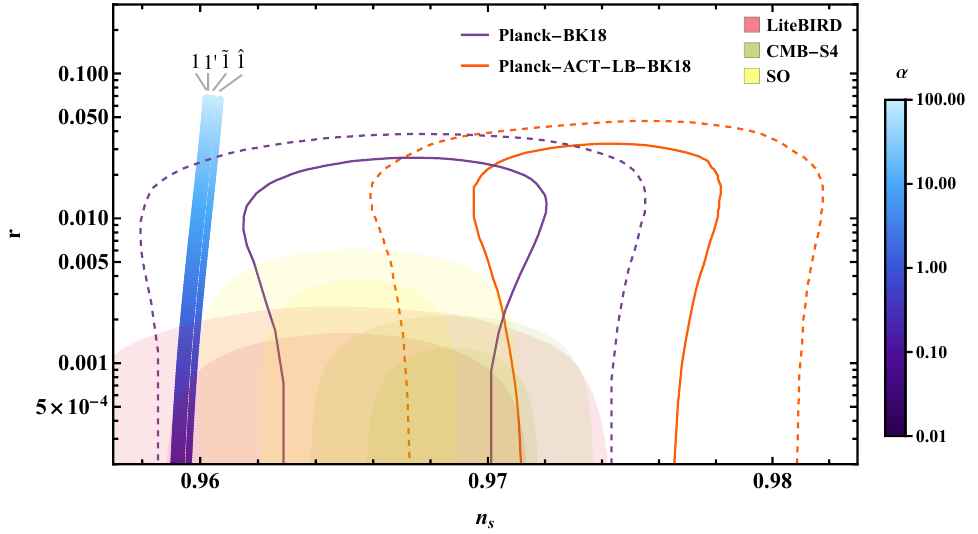}
        \caption{Tensor-to-scalar ratio $r$ vs. primordial tilt $n_s$ plot and for E-model in \Cref{tab:models} for quark model 1}
    \end{subfigure}%
    ~ \\
    \begin{subfigure}[t]{\textwidth}
        \centering
        \includegraphics{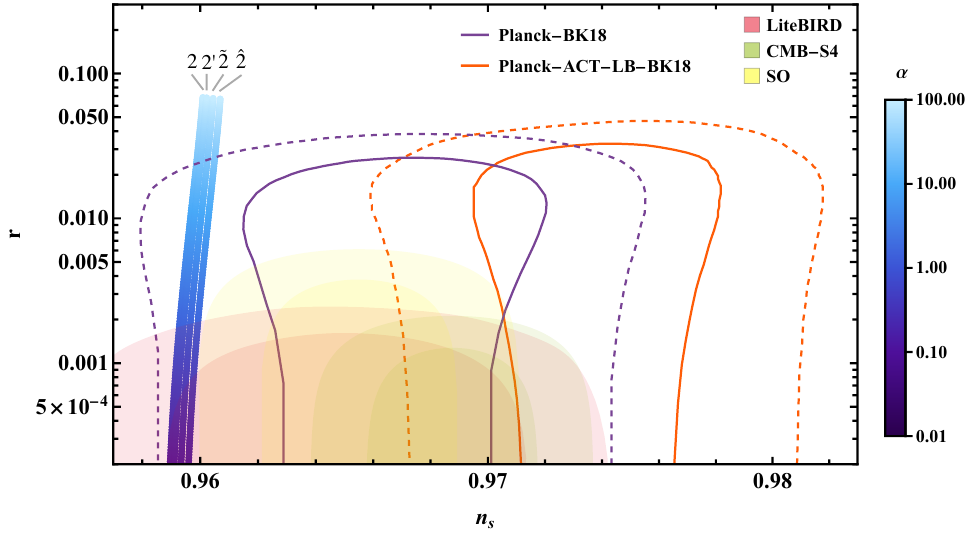}
        \caption{Tensor-to-scalar ratio $r$ vs. primordial tilt $n_s$ plot and for E-model in \Cref{tab:models} for quark model 2}
    \end{subfigure}
    \caption{\label{fig:nsrPlotTmodel}Results for T-model. The solid and dashed lines are the $1-$sigma and $2-$ sigma current constraints from \textit{Planck}, \textit{BK15}, and \textit{ACT} ~\cite{Planck:2018vyg,BICEP2:2018kqh}. The light and dark shaded regions represent future experimental constraints from \textit{LiteBIRD}, \textit{CMB-S4}, and \textit{SO}\cite{LiteBIRD:2022cnt,CMB-S4:2016ple,SimonsObservatory:2018koc}.}
\end{figure*}
\begin{figure*}[t!]\ContinuedFloat
    \centering
    \begin{subfigure}[t]{\textwidth}
        \centering
        \includegraphics{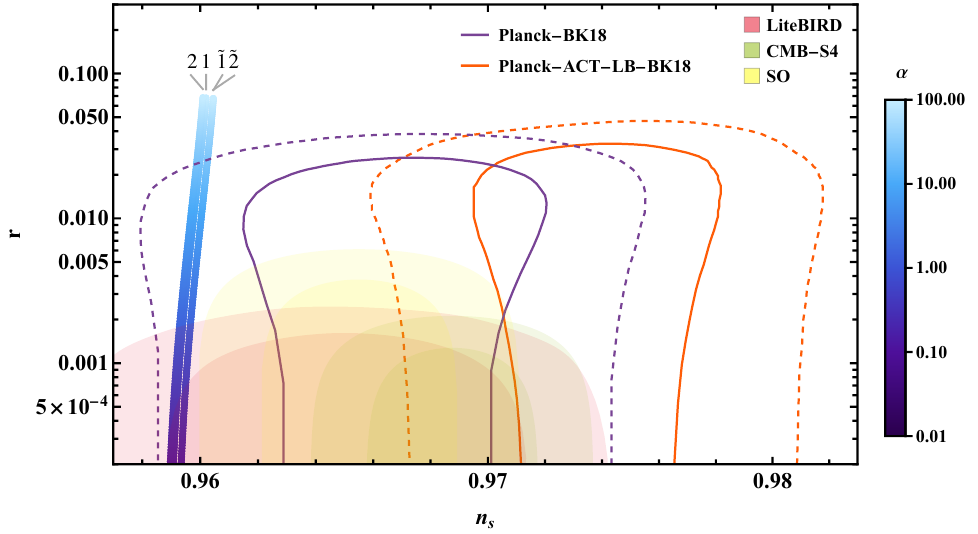}
        \caption{\label{fig:nsrPlotTmodelc}Tensor-to-scalar ratio $r$ vs. primordial tilt $n_s$ plot and for E-model in \Cref{tab:models} for model 1, 2(NO) and $\tilde{1}$, $\tilde{2}$(IO)}
    \end{subfigure}
    ~ \\
    \begin{subfigure}[t]{\textwidth}
        \centering
        \includegraphics{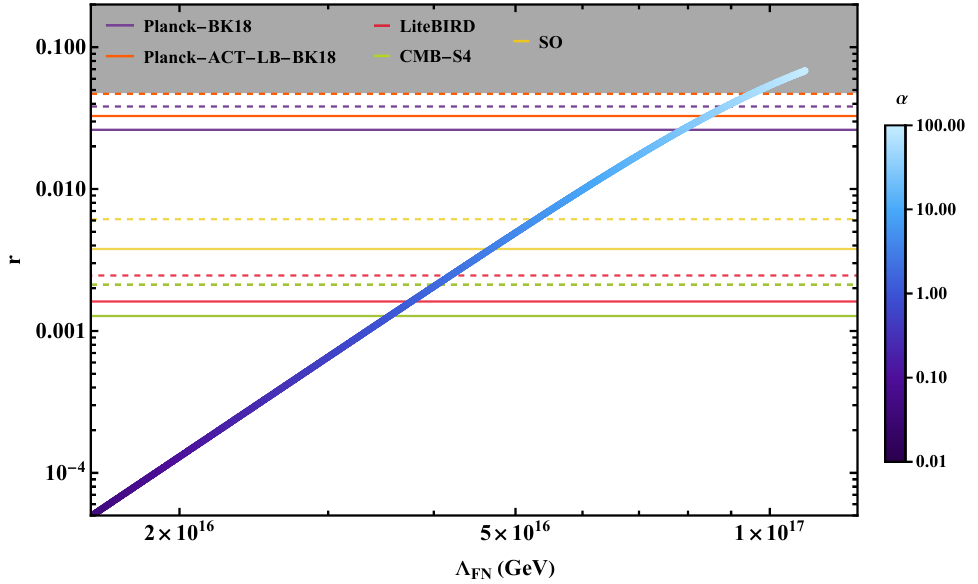}
        \caption{$r$ vs. \ac{FN} scale $\Lambda_{\FN}$ plot for T-model.}
    \end{subfigure}
    \caption{Results for T-model. The solid and dashed lines are the $1-$sigma and $2-$ sigma current constraints from \textit{Planck}, \textit{BK15}, and \textit{ACT} ~\cite{Planck:2018vyg,BICEP2:2018kqh}. The light and dark shaded regions represent future experimental constraints from \textit{LiteBIRD}, \textit{CMB-S4}, and \textit{SO}\cite{LiteBIRD:2022cnt,CMB-S4:2016ple,SimonsObservatory:2018koc}.}
\end{figure*}

\begin{figure*}[t!]
    \centering
    \begin{subfigure}[t]{\textwidth}
        \centering
        \includegraphics{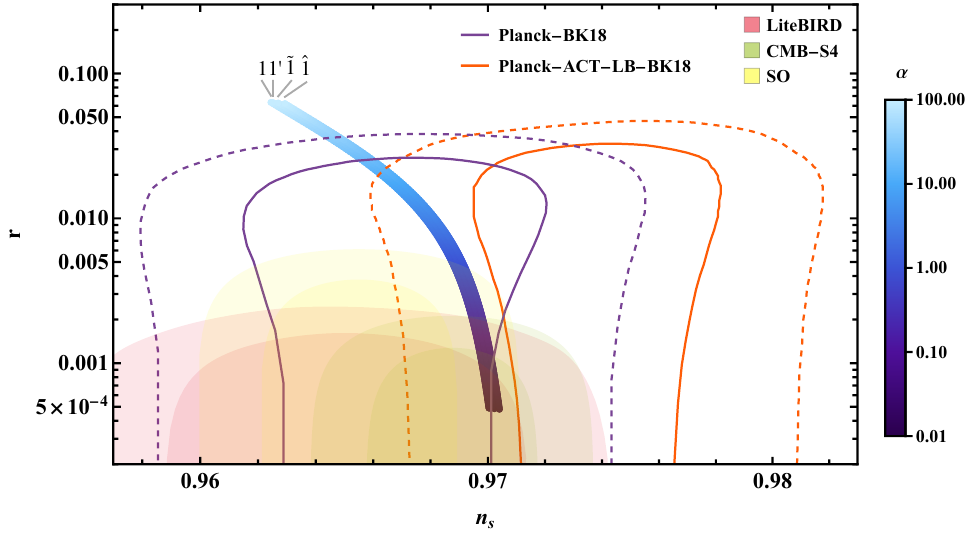}
        \caption{Tensor-to-scalar ratio $r$ vs. primordial tilt $n_s$ plot and for E-model in \Cref{tab:models} for quark model 1}
    \end{subfigure}%
    ~ \\
    \begin{subfigure}[t]{\textwidth}
        \centering
        \includegraphics{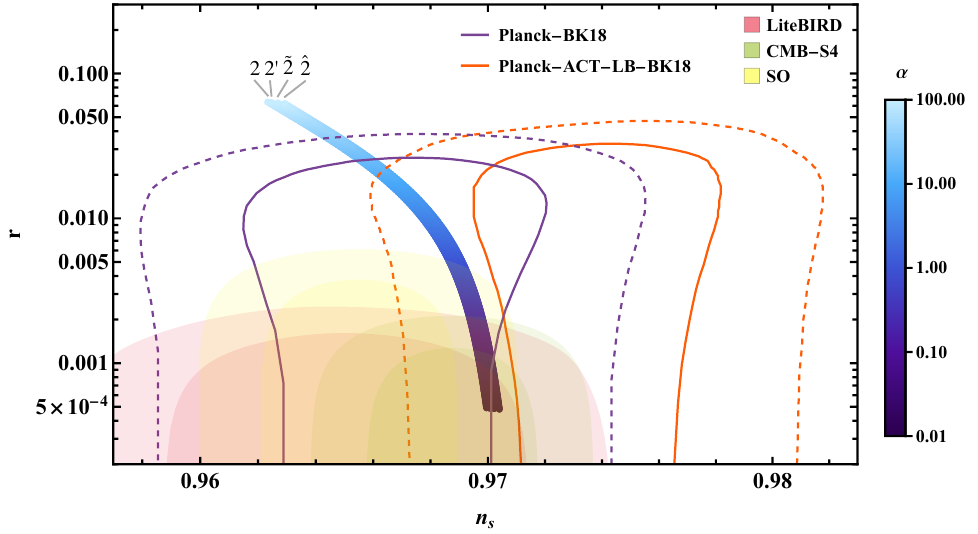}
        \caption{Tensor-to-scalar ratio $r$ vs. primordial tilt $n_s$ plot and for E-model in \Cref{tab:models} for quark model 2}
    \end{subfigure}
    \caption{\label{fig:nsrPlotPmodel}Results for P-model. The solid and dashed lines are the $1-$sigma and $2-$ sigma current constraints from \textit{Planck}, \textit{BK15}, and \textit{ACT} ~\cite{Planck:2018vyg,BICEP2:2018kqh}. The light and dark shaded regions represent future experimental constraints from \textit{LiteBIRD}, \textit{CMB-S4}, and \textit{SO}\cite{LiteBIRD:2022cnt,CMB-S4:2016ple,SimonsObservatory:2018koc}.}
\end{figure*}
\begin{figure*}[t!]\ContinuedFloat
    \centering
    \begin{subfigure}[t]{\textwidth}
        \centering
        \includegraphics{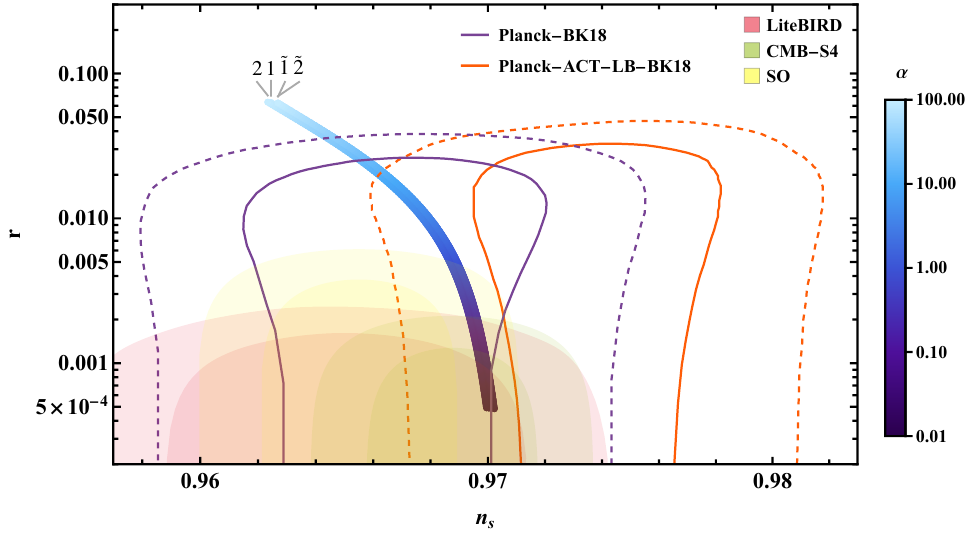}
        \caption{Tensor-to-scalar ratio $r$ vs. primordial tilt $n_s$ plot and for E-model in \Cref{tab:models} for model 1, 2(NO) and $\tilde{1}$, $\tilde{2}$(IO)}
    \end{subfigure}
    ~ \\
    \begin{subfigure}[t]{\textwidth}
        \centering
        \includegraphics{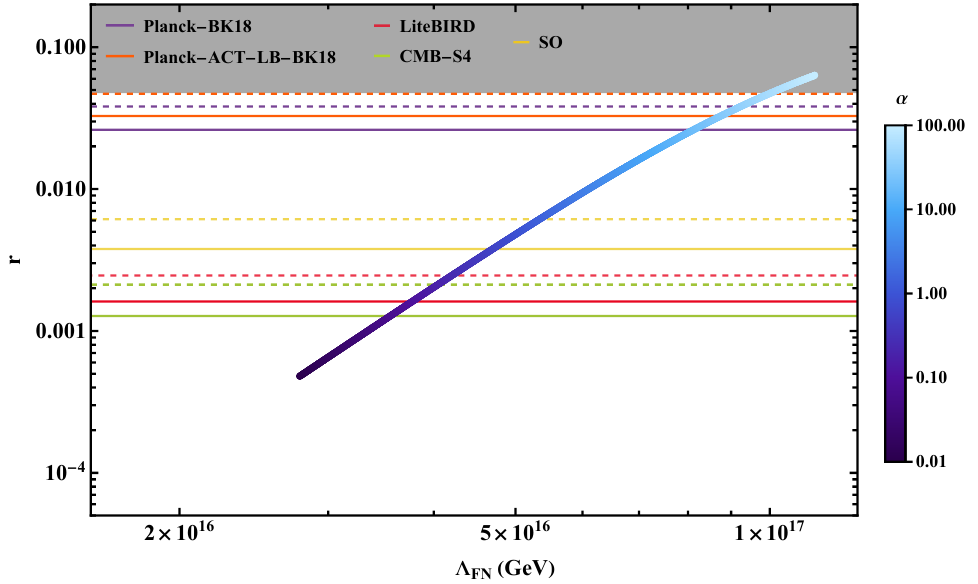}
        \caption{$r$ vs. \ac{FN} scale $\Lambda_{\FN}$ plot for P-model.}
    \end{subfigure}
    \caption{Results for P-model. The solid and dashed lines are the $1-$sigma and $2-$ sigma current constraints from \textit{Planck}, \textit{BK15}, and \textit{ACT} ~\cite{Planck:2018vyg,BICEP2:2018kqh}. The light and dark shaded regions represent future experimental constraint sensitivities from \textit{LiteBIRD}, \textit{CMB-S4}, and \textit{SO}\cite{LiteBIRD:2022cnt,CMB-S4:2016ple,SimonsObservatory:2018koc}.}
\end{figure*}

From \cref{eq:chiSM} we see that after the Higgs develops a \ac{VEV} $v_H$, then we can also have $2-$body decays. However, this decay is only viable if the reheating temperature is less than the electroweak symmetry breaking crossover temperature, approximately $\sim150$ GeV~\cite{DOnofrio:2015gop}. As we will see, for all physical cases in this work, the reheating temperature is much bigger than the electroweak symmetry breaking crossover temperature.\footnote{Nevertheless, also note that the three-body decay scales like $\nicefrac{g_{ij}^2 m_\chi^3}{\Lambda_{\fl}^2}$, while the two-body decay scales like $\nicefrac{g_{ij}^2 m_\chi v_\phi^2}{\Lambda_{\fl}^2}$. Then, the two-body decay channel is suppressed relative to the three-body decay channel by a factor of $\nicefrac{v_h^2}{m_\chi^2}\ll 1$. Hence, the inflaton decay width is always dominated by the three-body decay channel.} Thus, we ignore $2-$body decays in our analysis.

Before presenting the results of the representative models in \Cref{tab:models}, we mention that the they are selected to ensure that the contributions from the quark and lepton sectors are comparable.  In general, the dominant sector depends on the size of the couplings $g_{ij}^{a}$ in \cref{eq:gij} for each term. While most couplings are fixed by the experimental values of the fermion masses in each sector, the mass scale $\Lambda_\nu$ associated with the right-handed neutrinos remains a free parameter. A larger $\Lambda_\nu$ leads to a greater contribution from the lepton sector. This follows from the fact that the \ac{SM} neutrino masses, arising from the seesaw mechanism in \cref{eq:SM} are proportional to $\sim (g^{D} v)^2 / \Lambda_\nu $. Thus, to maintain the neutrino masses fixed, an increase in $\Lambda_\nu$ demands a corresponding increase in $g_{ij}^{D}$ (see \cref{eq:YukawaCouplings,eq:gij}). Conversely, a smaller $\Lambda_\nu$ leads to a dominant contribution from the quark sector to the inflavon decay width. We choose $\Lambda_\nu = 10^{7}$ GeV. This choice numerically ensures that the lepton and the quark sectors have both meaningful impact in the inflavon decay width. At the same time, it maintains perturbativity by demanding all Yukawa couplings $g_{ij}^{D}$ remain below $4\pi$.

We solve \cref{eq:GammaKey} using the decay width in \cref{eq:DecayWidthThreeBody} for the different \ac{FN} charge assignments in \Cref{tab:models}. The results are presented in \Cref{fig:nsrPlotEmodel,fig:nsrPlotTmodel,fig:nsrPlotPmodel} for the $E-$model, $T-$model and $P-$model respectively from \cref{eq:alpha-attractors}. Let us clarify the notation used for the models displayed in the Figures. For instance, the line labeled $1$ corresponds to a model where quarks are assigned to model $1$ and leptons to model $\circ$, as defined in \Cref{tab:models}. Similarly, the line labeled $\tilde{2}$ refers to a model with quarks in model $2$ and leptons in model $\tilde{\circ}$. In the top panels and the bottom-left panel, we show the scalar-to-tensor ratio $r$ as a function of the spectral index $n_s$ for different $\alpha$ values. Additionally, we also include the current constraints from \textit{Planck}, \textit{BK15}, \textit{BAO}~\cite{Planck:2018vyg,BICEP2:2018kqh} and ACT\cite{ACT:2025tim}, as well as experimental constraint forecasts from \textit{LiteBIRD}, \textit{CMB-S4}, and \textit{SO}\cite{LiteBIRD:2022cnt,CMB-S4:2016ple,SimonsObservatory:2018koc}.
In the last panel of \Cref{fig:nsrPlotEmodel,fig:nsrPlotTmodel,fig:nsrPlotPmodel}, we also plot the relation between the scale $\Lambda_\FN$ and the scalar-to-tensor ratio $r$. 

Notably, the observational bounds from ACT exclude our predictions for the $E$ and $T$ models. This results from the fact that the inclusion of the new ACT data leads to a  significant shift in the the value of $n_s$ compared to previous datasets. For instance, in~\cite{Haque:2025uri}, the predictions by the $E$ and $T$ models for different equation of states were compared against the addition of the new ACT data.  In their scenario, the reheating temperature is a free parameter. They determine the $T$ model with $n=1$ to be incompatible with the inclusion of the new ACT data, whereas the $E$ model with $n = 1$ can still be accommodated within the ACT observational bounds.  In contrast, in our scenario, the reheating temperature depends on the decay width, which is fixed by the \ac{FN} flavor model charges. As a result, one must solve \cref{eq:GammaKey} to compute the value of $n_s$. For our \ac{FN} scenario, an inflationary potential with the form of the $E$ and $T$ model and a matter-like equation of state $\overline{w}_{\re} = 0$, is completely ruled out by the combined data \textit{Planck}+\textit{ACT}+\textit{LB}+\textit{BK18}.

From \Cref{fig:nsrPlotEmodel,fig:nsrPlotTmodel,fig:nsrPlotPmodel}, we see that a larger $\alpha$ leads to a larger tensor-to-scalar ratio $r$. Most models lie within the bounds of current \textit{Planck}, \textit{BK15} and  \textit{BAO}, while values of $\alpha$ smaller than 1 could be ruled out in future experiments. Notably, the observational bounds from ACT exclude our predictions for the $E$ and $T$ models. For a fixed inflaton potential, the different \ac{FN} curves yield roughly the same form. However, note that values of $\alpha\gtrsim 10$ are ruled out by current values of observational bounds. On the other hand, very small values of $\alpha \lesssim 0.1$ could be ruled out by \textit{SO} forecast for some models.

\ac{IO} tends to yield larger decay width since, in general, it gives a greater minimum sum of neutrino masses compared to \ac{NO}. However, this is not necessarily true for individual models. Therefore, while cosmological data can constrain the decay width of the \ac{FN} model, it does not necessarily determine the mass ordering. 

In the last panels of \Cref{fig:nsrPlotEmodel,fig:nsrPlotTmodel,fig:nsrPlotPmodel}, we see that relationship between $\Lambda_\FN$ and the cosmological parameters $(n_s, r)$ is independent of the choice of \ac{FN} charges for a fixed $\alpha$. This is because $\Lambda_\FN$ enters only through the dimensionless parameter $\lambda = \frac{v_\phi}{\Lambda_\FN}$; thus, the decay width depends only on the ratio $\nicefrac{v_\phi}{\Lambda_\FN}$, which is fixed to 0.17 in this work, causing the decay width to be insensitive to $\Lambda_{\FN}$.

In \Cref{fig:gamma-rVns}, we show the relation between the scalar-to-tensor ratio $r$ with the spectral index $n_s$ for a fixed value of $\alpha = 0.1$ while varying the value of the inflaton decay width $\Gamma$. 
We find that larger decay widths correspond to larger values of $n_s$ for $n=1$. This relation between the decay width and $n_s$ is also observed for the $T$ and $P$ models for different values of $\alpha$. The vertical red dotted-dashed line indicates the region where $N_{\re} = 0$. For $n=1$, the left-side of this line denotes the physical scenarios where $N_{\re} > 0$. Finally, in \Cref{app:SuppReheatingPlot} we show that for $n=1$ the reheating temperature lies in the parametric window $ T_{\mathrm{EW}} \ll T_{\re} \ll M_{\pl}$, where $T_{\mathrm{EW}} = 150~\mathrm{GeV}$ is the electroweak crossover temperature. In this regime, the three-body decay channels in \Cref{eq:DecayWidthThreeBody} remain kinematically allowed, while the two-body decay is forbidden. Hence, the reheating temperature lies within consistent bounds.

\subsection{$\alpha-$attractor potential for $n\neq 1$.}
\label{subsec:alphanNeq1}

\begin{figure}
    \centering
    \begin{subfigure}[t]{\textwidth}
        \centering
        \includegraphics{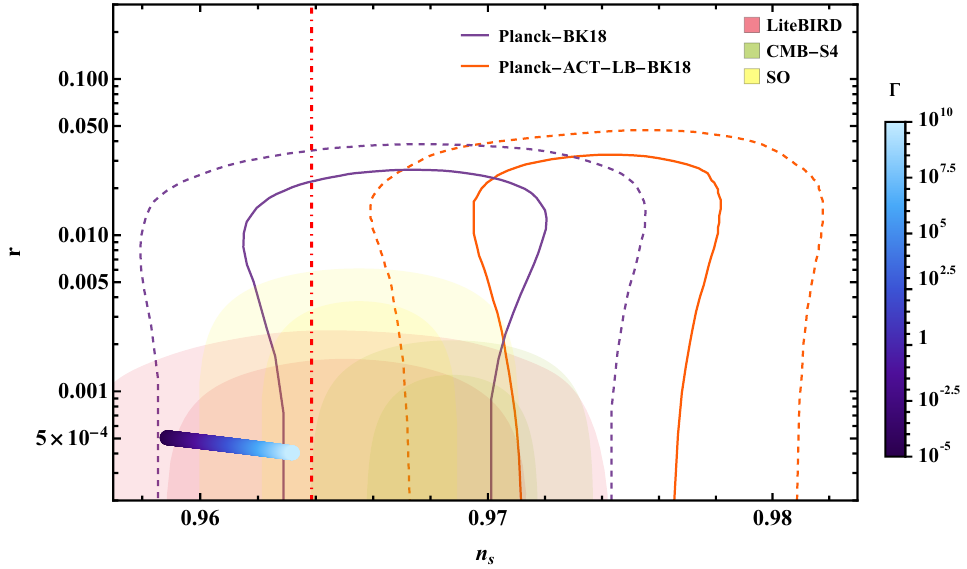}
        \caption{$n = 1$ case}
    \end{subfigure}%
    ~ \\
    \begin{subfigure}[t]{\textwidth}
        \centering
        \includegraphics{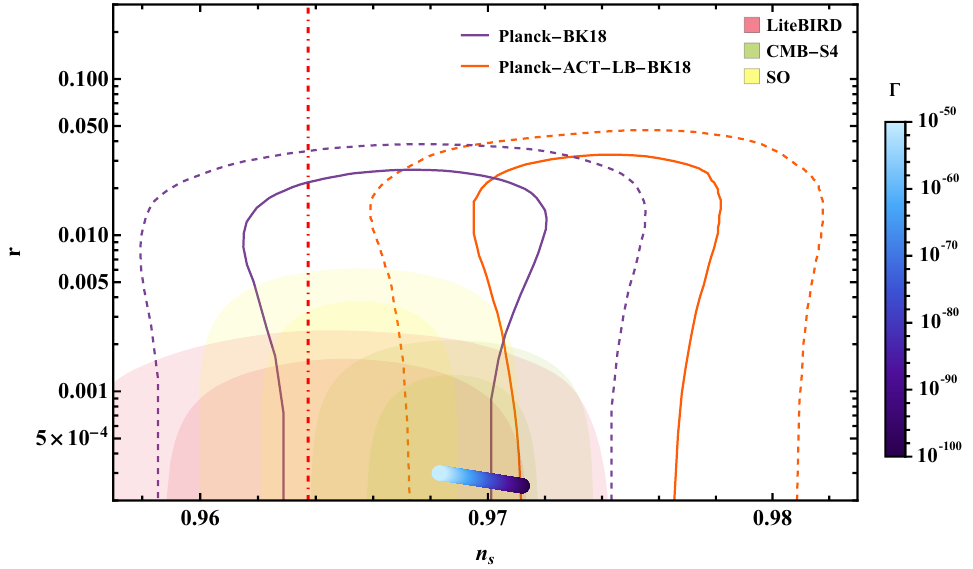}
        \caption{$n = 3$ case}
    \end{subfigure}
    \caption{\label{fig:gamma-rVns}The scalar-to-tensor ratio $r$ as a function of the spectral index $n_s$ for a fixed $\alpha = 0.1$ in the $E$-model. The color bar indicates different values of the inflaton decay width $\Gamma$. The red dotted-dashed line indicates when $N_{\re} = 0$. The physical region $(N_{\re} > 0)$ corresponds to the area to the left(right) of the red dotted-dashed line for $n=1$ ($n=3$). }
\end{figure}

For $n\neq 1$ in \cref{eq:alpha-attractors}, the classical inflaton mass vanishes (see \cref{eq:mchi2}). We then instead compute the perturbative decay of the oscillating inflaton, which acquires an effective mass and an effective coupling. 
To compute the decay width for $n\neq 1$, we follow \cite{Garcia:2020wiy, Chakraborty:2023ocr}. In this framework, the scalar field is approximated as $\phi(t) = \phi_0(t) \mathcal P(t)$. The function $\mathcal P(t)$ is quasi-periodic and captures the information of short time-scale oscillations. In contrast, the function $\phi_0 (t)$ carries the effect of long time-scale oscillations, which is approximated as a constant. With this approximation, the decay width is given by
\begin{equation}
\label{eq:DecayWidtEff}
        \left(\Gamma_{\left( n \neq 1 \right)}\right)_{ij}^{a} ~ = ~\frac{N_c}{3} \frac{\left|(g^{\text{eff}})_{ij}^{a}\right|^2}{64 \pi^3}\left|\frac{v_\phi}{2 \Lambda_\fl} \sqrt{\frac{2}{3\alpha}} \right|^2 (m_\chi^{\text{eff}})^3\;
        \text{, where }\; 
        m_\chi^{\text{eff}} = \frac{\partial^2 V(\phi_0(t))}{\partial \phi_0^2}\bigg|_{t=t_{\re}}\;.
\end{equation}
Here the effective mass $m_\chi^{\text{eff}}$ is computed at the end of reheating $t_{\re}$. The effective coupling $g^{\text{eff}}$ is given by
\begin{align}
\label{eq:geff}
    g^{\text{eff}} =& 
        g \sqrt{\left(\frac{\omega}{m_\chi^{\text{eff}}}\right)^3 (2n+2)(2n-1)\sum_\ell |\mathcal P_\ell|^2 \ell^3 
        }\;,
        \\
        &\text{where }\; 
        \omega = m_\chi^{\text{eff}} \sqrt{\frac{\pi n}{2n-1}}\frac{\mathsf \Gamma(\frac{1}{2}+\frac{1}{2n})}{\frac{1}{2n}}\;,\nonumber
\end{align}
where we have used the fact that \(m_\chi^{\text{eff}}\) is much larger than the masses of other particles. Here $\mathsf \Gamma$ is the gamma function. In \Cref{fig:gamma-rVns}, we show the relation between $r$ and $n_s$ for $n=3$, $\alpha = 0.1$ and different values of the decay width $\Gamma$. We display a broad range of $\Gamma$ values. In \Cref{app:SuppReheatingPlot}, we show that for all models in \Cref{tab:models}, and for each of the three inflationary scenarios ($E$-, $T$-, and $P$-models), the reheating temperature remains below the \ac{BBN} temperature of $\mathcal{O}(1\,\mathrm{MeV})$. Therefore, inflavon \ac{FN} models with $\neq 1$ are incompatible with \ac{BBN} constraints and are not phenomenologically viable.

\section{Discussion and Conclusion}
\label{sec:Conclusion}

We presented a scenario where the scalar flavon breaking the global $\U{1}_{\FN}$, also drives cosmic inflation in early universe. The \ac{FN} models considered here include both quarks and leptons, where the neutrinos get their mass through a type-1 seesaw mechanism with three right-handed neutrinos. We show that the flavon, \textit{dubbed as the inflavon}, generates an $\alpha-$attractor potential from a non-canonical inflavon kinetic term. We consider three types of $\alpha-$attractor models: the $E-$model, the $T-$model and the polynomial or $P-$model. The inflavon, then, reheats the visible Universe by transferring its energy into particles of the \ac{SM} or right-handed neutrinos through the Dirac mass terms. Assuming that the right-handed neutrinos are not charged under the \ac{FN} group, the main decay channel is a three-body inflavon decay. This connection between flavor physics and inflation is possible by using the relation in \cref{eq:GammaKey}, which dictates that the inflavon decay width matches the Hubble rate at the end of reheating. Since the inflavon decay width is determined by the yukawa couplings and the \ac{FN} charges, then these quantities can be related to inflationary observables which determine the Hubble rate, depending on the inflation potential.

By choosing one of the three $\alpha-$attractor potentials and a set of \ac{FN} charge assignments, we can obtain two \ac{CMB} observables: the spectral index $n_s$ and the scalar-to-tensor ratio $r$ as a function of the value $\alpha$. We calculated these \ac{CMB} observables for eight different sets of \ac{FN} charge assignments that do not exhibit fine-tuning. We expect our results to not differ from the eight models considered for different other sets of \ac{FN} charges. Furthermore, we fix the neutrino Majorana mass scale $\Lambda_\nu$ to $10^{7}$ GeV. With this value, the inflavon decay widths into the lepton and quark sectors are of comparable magnitude while remaining within the perturbative regime. For an $\alpha-$attractor potential with a non-vanishing classical mass, we calculated the predictions of $n_s$ and $r$ for several values of $\alpha$. Furthermore, we compared our predictions with the current experimental bounds from \textit{Planck}, \textit{BK15}, \textit{BAO} and \textit{ACT} and the forecasts from \textit{LiteBIRD}, \textit{CMB-S4} and \textit{SO}. Remarkably, the bounds from \textit{ACT} completely exclude our predictions for \ac{FN} models with $E$ and $T$ models. We do not observe any differences between models with normal and inverted neutrino mass ordering. Furthermore, even though different \ac{FN} charge assignments yield different predictions for $n_s$ and $r$, the different predictions vary only slightly. However, future experiments distinguish between and potentially rule out various \ac{FN} models. For instance, as illustrated in \Cref{fig:nsrPlotTmodelc}, the \textit{SO} forecast could exclude a particular \ac{FN} model. However, current and future bounds can rule out a range of values for $\alpha$ given a specific \ac{FN} model. We also determined the predicted \ac{FN} scale within this framework. Its value lies in the range of $10^{16}$–$10^{17}$ GeV, depending on the value of $\alpha$. Finally, we did the same exercise of calculating the predictions of $n_s$ and $r$ for an $\alpha-$attractor potential with vanishing classical mass, using the eight \ac{FN} charge assignments considered in this work. In this case, the decay width of the inflavon is highly suppressed, yielding an unphysical reheating temperature below the \ac{BBN} temperature of around $1\,$MeV. This behavior is summarized in \Cref{app:SuppReheatingPlot} where we plot the reheating temperature as a function of the spectral index. We see that the reheating temperature is significantly suppressed for the case of a vanishing classical mass.

We showed that \ac{CMB} observables can not only constrain the parameter space of \ac{FN} flavor models when embedded within an inflationary framework, but that future \ac{CMB} measurements will also be able to probe a broad range of \ac{FN} scenarios. Obviously, the train of thought in this article can be extended beyond the \ac{FN} flavor symmetry to other types of flavor symmetries, such as discrete flavor symmetries and modular flavor symmetries. Future directions include considering alternative inflaton potentials, introducing additional inflavon fields, and exploring other cosmological and astrophysical observables, such as primordial gravitational waves. Moreover, as future \ac{CMB} experiments achieve greater precision, it may become possible to constrain flavor models through the running of the spectral index or even running of the running. Moreover, if such flavons decay to dark sector like flavon-portal dark matter \cite{Barman:2021yaz} or baryogenesis or leptogenesis \cite{Borboruah:2024eal}, such non-thermal particle production can be tested via CMB. We leave these intriguing questions for future work.

\section*{Acknowledgement}
We thank Kenji Nishiwaki for discussion.
The work of M.-C.C. and X.-G.L. was partially supported by U.S. National Science Foundation (NSF) under grant number PHY-2210283. C.M.-S. acknowledges the support of the US NSF Graduate Research Fellowship Program (GRFP) and the Eugene Cota-Robles Fellowship provided by the University of California, Irvine. V.K-P. acknowledges the support of the Miguel Velez Scholarship provided by the University of California, Irvine. M.-C.C. and A.G. wish to acknowledge the BCVSPIN 2023 School (Kathmandu, Nepal) along with the hospitality of Kathmandu and Tribhuvan Universities for providing the stimulating and collaborative environment that facilitated the development of this project \href{https://www.bcvspin.org/}{BCVSPIN organization}, \href{https://indico.cern.ch/event/1253238/overview}{BCVSPIN 2023 Nepal}.

\appendix

\section{Higgs-Flavon mixing}
\label{app:Higgs-Flavon-Mixing}
As noted in the main text, the Lagrangian term given by
\begin{equation}
\label{eq:phih-mixing}
    \mathcal{L}_{\mathrm{\phi-h}} ~=~ \lambda_{\phi H} |\phi|^2 |H|^2\;,
\end{equation}
is unavoidable. Since we want that the main contribution to the inflaton decay comes from the \ac{FN} mechanism in \cref{eq:BeforeFlavorL}, then we estimate how small the $\lambda$ coupling should. We first proceed to compute an estimate of the decay width of the inflaton three body decay given by \cref{eq:chiSM} using \cref{eq:gij,eq:DecayWidthThreeBody}. Note that $\alpha_{ij},n_{ij}\lesssim\mathcal{O}(1)$ and $\lambda^{|n_{ij}|}<1$. Then, we have
\begin{equation}
\label{eq:ThreebodyBound_1}
    \Gamma_{ij}^{a} ~\lesssim~ \frac{1}{64\pi^3 \alpha}\frac{m_\chi^3}{\Lambda_{\fl}^2}\;.
\end{equation}
Then, we fix $\Lambda_\fl = M_{\pl}$ as in \Cref{sec:FNFaceCosmology}. Furthermore, from the numerical evaluations performed for 
\Cref{fig:nsrPlotEmodel,fig:nsrPlotTmodel,fig:nsrPlotPmodel,app:AlphaAtractors} we find that the inflavon mass is of order $m_\chi \sim 10^{13}\,$GeV. From \cref{eq:ThreebodyBound_1}, we get that
\begin{equation}
    \label{eq:ThreebodyBound_2}
    \Gamma_{ij}^{a} ~\lesssim~ \frac{0.1}{\alpha}\;\mathrm{GeV}\;.
\end{equation}
Note that if the maximum value of $\lambda^{|n_{ij}^{a}|}$ for some $n_{ij}^{a}$ in a given \ac{FN} model is much smaller than $0.1$. We now compute the decay width for the Higgs pair channel from \cref{eq:phih-mixing}. Using \cref{eq:flavonVev,eq:SigmaAndChi} in \cref{eq:ThreebodyBound_2} gives
\begin{equation}
\label{eq:phih-mixing2}
    \mathcal{L}_{\mathrm{\phi-h}} ~\supset~ \sqrt{\frac{2}{3\alpha}}\lambda_{\phi H} \frac{v_\phi^2}{\Lambda_\fl}\sigma|H|^2\;.
\end{equation}
Thus, the decay width is of order
\begin{equation}
\label{eq:HiggsDecay}
    \Gamma_{\chi\to HH} ~\sim~ \left| \lambda_{\phi H} \sqrt{\frac{2}{3\alpha}}\frac{v_\phi^2}{\Lambda_\fl}\right|^2 \frac{1}{m_\chi}\;.
\end{equation}
Using \cref{eq:mchi2} with $v_\phi / \Lambda_{\FN} \sim 0.1$, $\Lambda_{\inf} = 0.1 \Lambda_{\FN}$ and $\Lambda_\fl = M_{\pl }$ in \cref{eq:HiggsDecay} yields
\begin{equation}
\label{eq:GammachiHH}
    \Gamma_{\chi\to H H}~\sim~ \lambda_{\phi H}^2 \frac{\Lambda_\FN^2}{\sqrt{\alpha}\,10^2\,M_\pl}\;.
\end{equation}
Thus, demanding that $\Gamma_{\chi\to HH} \ll \Gamma_{ij}^a$ gives
\begin{equation}
    \lambda_{\phi H}^2~\ll~ \frac{\left(10\;\mathrm{GeV}\right) M_{\pl}}{\sqrt{\alpha}\Lambda_{\FN}^2}\;.
\end{equation}
From the numerical evaluations in \Cref{fig:nsrPlotEmodel,fig:nsrPlotTmodel}, we find that the smallest value of $\left(\sqrt{\alpha} \Lambda_{\FN}^2\right)^{-1}$ considered in this work is around $10^{-36}$ GeV$^{-1}$. Thus, we find that 
\begin{equation}
\label{eq:FinalLambdaPhiHBound}
    \lambda_{\phi H} \ll 10^{-9}\;.
\end{equation}

In principle, the bound found in \cref{eq:FinalLambdaPhiHBound} could be corrected if loop effects become important. However, note that the physical reheating temperature considered in this work is much bigger than the electroweak crossover transition temperature. This means that all couplings between fermions, the Higgs and the inflaton are four body couplings. Thus, loop processes that could change the decay widths $\Gamma_{ij}^{(a)}$ and $\Gamma_{\chi\to H H}$ have more than one loop. On the other hand, for the coupling $\lambda$ there are $1-$loop corrections due to Higgs self-energies. Since the Higgs quartic coupling $\lambda_H$ is smaller than one, these corrections are subdominant. Hence, our bound in \cref{eq:FinalLambdaPhiHBound} should hold at the loop-level.

\section{$\alpha-$attractor potentials from $\U{1}_\FN$ flavon potential}
\label{app:AlphaAtractors}
In this Appendix, we will write the explicit formulas we use in \Cref{sec:CMB-alpha} for the three $\alpha-$attractor potentials we consider: E-Model, T-Model and P-Model. Furthermore, unlike in \Cref{sec:CMB-alpha}, we will omit the explicit arguments of the functions, as they should be clear from context. We start with the $\alpha-$attractor potentials from \cref{eq:alpha-attractors} and we rewrite them here for completeness
\begin{equation}
    V ~:=~ \begin{cases}
        \Lambda_{\mathrm{inf}}^4\left(1-\exp\left[-\sqrt{\frac{2}{3\alpha}}\frac{\chi}{\Lambda_{\fl}}{}\right]\right)^{2n}\;,\quad\text{E-Model}\;,\\        \Lambda_{\mathrm{inf}}^4\left(\tanh\left[ \sqrt{\frac{2}{3\alpha}}\frac{\chi}{\Lambda_\fl}\right]\right)^{2n}\;,\quad\text{T-Model}\;,  \\ 
         \Lambda_{\mathrm{inf}}^4\frac{\chi^{2n}}{\chi^{2n}+\left( \sqrt{\frac{3\alpha}{2}}\Lambda_\fl\right)^{2n}}\;,\quad\text{P-Model}\;.
    \end{cases}
    \label{eq:alpha-attractors-appendix}
\end{equation}

The slow-roll parameters computed from \cref{eq:SlowRoll} are given by
\begin{equation}
\label{eq:epsilon-appendix}
    \epsilon ~=~ \begin{cases}
            \frac{4 n^2M_{\pl}^2}{
3  \alpha \Lambda_{\fl}^2
}\left( 1 - \ee^{\sqrt{\frac{2}{3\alpha}}\frac{ \chi}{\Lambda_{\fl}}} \right)^{-2}\;,\quad\text{E-Model}\;,\\
\frac{
16  n^2 M_{\pl}^2 
}{3  \alpha \Lambda_{\fl}^2}\mathrm{csch}^2 \left(2\sqrt{\frac{2}{3\alpha }} \frac{\chi}{\Lambda_{\fl}} \right)\;,\quad\text{T-Model}\;,\\
\frac{2M_{\pl}^2 9^n n^2\alpha^{2n} \Lambda_{\fl}^{4n}
}{\left( 3^n \alpha^n \Lambda_{\fl}^{2n} \chi + 2^n \chi^{2n+1} \right)^2}\;,\quad\text{P-Model}\;,
    \end{cases}
\end{equation}
\begin{equation}
\label{eq:eta-appendix}
    \eta ~=~ \begin{cases}
        \frac{
4  nM_{pl}^2
}{
3 \alpha \Lambda_{\fl}^2
}\left( 1 - \ee^{\sqrt{\frac{2}{3\alpha}}\frac{ \chi}{\Lambda_{\fl}}} \right)^{-2}\left( -2n + \ee^{\sqrt{\frac{2}{3\alpha}}\frac{ \chi}{\Lambda_{\fl}}} \right)\;,\quad\text{E-Model}\;,\\
\frac{16 n M_{\pl}^2  }{3 \alpha \Lambda_{\fl}^2
}\left[ 2n - \cosh \left( 2 \sqrt{\frac{2}{3\alpha}} \frac{\chi}{\Lambda_{\fl}} \right) \right]
\text{csch}^2 \left(2 \sqrt{\frac{2}{3\alpha}} \frac{\chi}{\Lambda_{\fl}} \right)\;,\quad\text{T-Model}\;,\\
\frac{2M_{\pl}^2 3^n n\alpha^n \Lambda_{\fl}^{2n} \left( 3^n(2n-1)\alpha^n \Lambda_{\fl}^{2n} - 2^n(2n+1) \chi^{2n} \right)}{\left( 3^n \alpha^n \Lambda_{\fl}^{2n} \chi + 2^n \chi^{2n+1} \right)^2}\;,\quad\text{P-Model}\;.
    \end{cases}
\end{equation}
Then, using \cref{eq:ns}, we get
\begin{equation}
\label{eq:ns-appendix}
    n_s ~=~ \begin{cases}
        1 - \frac{
8 n M_{\pl}^2
}{
3  \alpha \Lambda_{\fl}^2
}\left(n + \ee^{\sqrt{\frac{2}{3\alpha}}\frac{ \chi_k}{\Lambda_{\fl}}}  \right)\left( 1 - \ee^{\sqrt{\frac{2}{3\alpha}}\frac{ \chi_k}{\Lambda_{\fl}}}  \right)^{-2}\;\quad\text{E-Model}\;,\\
1 - \frac{8 n  M_{\pl}^2 }{3 \alpha \Lambda_{\fl}^2
} \left(
(1 + n) \, \text{csch}^2\left( \sqrt{\frac{2}{3\alpha}}\frac{ \chi_k}{\Lambda_{\fl}} \right)
+ (1 - n) \, \text{sech}^2\left( \sqrt{\frac{2}{3\alpha}}\frac{ \chi_k}{\Lambda_{\fl}}\right)
\right)\;,\quad\text{T-Model}\;,\\
1 - \frac{4 M_{\pl}^2  3^n n \alpha^n  \left( 3^n(1+n)\alpha^n \Lambda_{\fl}^{4n} + 2^n(1+2n)\Lambda_{\fl}^{2n}\chi_k^{2n} \right)}{\left( 3^n \alpha^n \Lambda_{\fl}^{2n} \chi_k + 2^n \chi_k^{2n+1}\right)^2}\;,\quad\text{P-Model}\;,
    \end{cases}
\end{equation}
\begin{equation}
\label{eq:r-appendix}
    r ~=~ \begin{cases}
        \frac{64 n^2M_{\pl}^2}{
3  \alpha \Lambda_{\fl}^2
}\left( 1 - \ee^{\sqrt{\frac{2}{3\alpha}}\frac{ \chi_k}{\Lambda_{\fl}}} \right)^{-2}\;\quad\text{E-Model}\;,\\
\frac{
256 n^2  M_{\pl}^2 }{3 \alpha \Lambda_{\fl}^2} \text{csch}^2 \left( 2\sqrt{\frac{2}{3\alpha}}\frac{ \chi_k}{\Lambda_{\fl}}\right)\;,\quad\text{T-Model}\;,\\
\frac{2^5M_{\pl}^2 9^n n^2\alpha^{2n} \Lambda_{\fl}^{4n}
}{\left( 3^n \alpha^n \Lambda_{\fl}^{2n} \chi + 2^n \chi^{2n+1} \right)^2}\;,\quad\text{P-Model}\;,
    \end{cases}
\end{equation}
\begin{equation}
\label{eq:As-appendix}
    A_s ~=~ \begin{cases}
        \frac{
\alpha \Lambda_{\fl}^2 \, \Lambda_{\inf}^4 
}{
32 n^2M_{\pl}^6 \pi^2
}\left( 1 - \ee^{\sqrt{\frac{2}{3\alpha}}\frac{ \chi_k}{\Lambda_{\fl}}}  \right)^{2}\left( 1 - \ee^{-\sqrt{\frac{2}{3\alpha}}\frac{ \chi_k}{\Lambda_{\fl}}}  \right)^{2n}\;,\quad\text{E-Model}\;,\\
\frac{\alpha \, \Lambda_{\fl}^2 \, \Lambda_{\inf}^4}{
128 n^2 M_{\pl}^6 \pi^2
}\sinh^2 \left( 2 \sqrt{\frac{2}{3\alpha}}\frac{ \chi_k}{\Lambda_{\fl}} \right)
\tanh^{2n} \left( \sqrt{\frac{2}{3\alpha}}\frac{ \chi_k}{\Lambda_{\fl}}\right)\;,\quad\text{T-Model}\;,\\
\frac{\Lambda_{\inf}^4 \chi_k^{2+2n} \left( 3^n \alpha^n \Lambda_{\fl}^{2n} + 2^n \chi_k^{2n} \right)}{2^{4-n}3^{1+2n} M_{\pl}^6 \pi^2 n^2 \alpha^{2n} \Lambda_{\fl}^{4n}
}\;,\quad\text{P-Model} 
    \end{cases}
\end{equation}
Below, we uniformly fix $n=1$ for the P model in order to obtain the analytical expressions of many physical quantities.
First, From \cref{eq:ns-appendix} we can solve for $\chi_k$ from $n_s$ as follows
\begin{align}
\label{eq:chik-appendix}
    \chi_k ~=~ \begin{cases}
        \sqrt{\frac{3 \alpha}{2}} \Lambda_{\fl} \log \left[
1 + \frac{
2 M_{\pl} \left( 2 M_{\pl} n + \sqrt{4 M_{\pl}^2 n^2 + 6 n (1 + n)(1 - n_s) \alpha \Lambda_{\fl}^2} \right)
}{
3 (1 - n_s) \alpha \Lambda_{\fl}^2
}
\right]\;,\quad\text{E-Model}\;,\\
\frac{1}{2} \sqrt{\frac{3 \alpha}{2}} \Lambda_{\fl}\log \left[\frac{\theta_3 + 16 _{\pl}^2n^2 + 4 M_{\pl}\sqrt{2 n \left( \theta_1 + \theta_3\right)}}{3\left( 1- n_s \right) \alpha \Lambda_{\fl}^2} \right]\;,\quad\text{T-Model}\\
   \sqrt{\frac{-2\alpha\Lambda_{\inf}^2 + \Theta}{2}}\;,\quad\text{P-Model \,($n=1$)}\;,
    \end{cases}
\end{align}
and \cref{eq:As} we can solve for $\Lambda_{\inf }$. We get
\begin{equation}
\label{eq:Lambdainf-appendix}
    \Lambda_{\inf } ~=~ \begin{cases}
        M_{\pl} \bigg(\frac{3 \pi^2  r A_s}{2}\bigg)^{1/4}\bigg[\frac{2 M_{\pl}\left(1+2n\right)+\sqrt{4 M_{\pl}^2 n^2 + 6\Lambda_{\inf}^2\alpha n(1+n)(1-n_s)}}{4 M_{\pl}n (1+n)}\bigg]^{n/2}\;,\quad\text{E-Model}\;,\\
        M_{\pl} \bigg(\frac{ \pi^2  r A_s}{6}\bigg)^{1/4}\bigg[\frac{12\sqrt{2}M_{\pl}\left(n-1\right)n\left(1-n_s\right)\alpha\Lambda_{\fl}^2}{\left( \theta_2 - \theta_3\right)\sqrt{ n\left(\theta_1 + \theta_3\right)}}\bigg]^{n/2}\;,\quad\text{T-Model}\;,\\
        M_{\pl}\left( \frac{3 \pi^2 r}{2}\right)^{1/4}\left[ \frac{\alpha\Lambda_{\fl}^2 + \Theta}{-2 \alpha \Lambda_{\fl}^2 +\Theta}\right]\;,\quad\text{P-Model \,($n=1$)}\,,
    \end{cases} 
\end{equation}
where we defined
\begin{align}
    \theta_1 & ~:=~ 16 M_{\pl}^2 n + 3n(1-n_s)\alpha\Lambda_{\inf}^2\;,\\
    \theta_2 & ~:=~ 16 M_{\pl}^2 n + 3(1-n_s)\alpha\Lambda_{\inf}^2\;,\\
    \theta_3 & ~:=~\sqrt{256 M_{\pl}^4 n^2 + 96 M_{\pl}^2 n^2\left(1 - n_s \right)\alpha\Lambda_{\fl}^2+9\left(1-n_s\right)^2\alpha^2\Lambda_{\fl}^4}
\end{align}
\begin{align}
    \Theta & ~:=~ \left(\alpha ^3 \Lambda_{\fl}^6 -6 \sqrt{2} \Delta\right)^{1/3}+\left(\alpha ^3 \Lambda_{\fl}^6+6 \sqrt{2} \Delta\right)^{1/3}\;,\\
    \Delta & ~:=~ \bigg[\frac{\alpha ^3 \Lambda_{\fl}^6 M_{\pl}^2 \bigg(192 M_{\pl}^4-24 \alpha  \Lambda_{\fl}^2 M_{\pl}^2 (n_s-1)+\alpha ^2\Lambda_{\fl}^4 (n_s-1)^2\bigg)}{(n_s-1)^3}\bigg]^{1/2}\; .
\end{align}
Furthermore, we can obtain the field value when inflation ends at $\epsilon = 1$ from \cref{eq:epsilon-appendix}, we get
\begin{equation}
\label{eq:chiend-appendix}
    \chi_\mathrm{end} ~=~ \begin{cases}
        \sqrt{\frac{3 \alpha}{2}} \Lambda_{\fl} \log \left[ 1+\frac{2 n M_{\pl}  }{\sqrt{3\alpha} \, \Lambda_{\fl}} \right]\;,\quad\text{E-Model}\;,\\
        \frac{\Lambda_{\fl}}{4} \sqrt{\frac{3\alpha}{2}} 
\log \left[
1 + \frac{8 n M_{\pl} \left( 4 n M_{\pl}  + \sqrt{16 M_{\pl}^2 n^2 + 3 \alpha \Lambda_{\fl}^2} \right)}{3 \alpha \Lambda_{\fl}^2}
\right]\;,\quad\text{T-Model}\;,\\
\frac{
    -\alpha \Lambda_{\fl}^2 + \left( \alpha^2 \Lambda_{\fl}^4 \Theta_2 \right)^{1/3}
}{
    \sqrt{2} \left( \alpha^2 \Lambda_{\fl}^4 \Theta_2 \right)^{1/6}
}\;,\quad\text{P-Model \,($n=1$)}\;.
    \end{cases}
\end{equation}
Inserting \cref{eq:chiend-appendix} into the $\alpha-$attractor potential of \cref{eq:alpha-attractors-appendix} yields the potential
\begin{equation}
\label{eq:Vchiend}
    V(\chi_\mathrm{end}) ~=~  \begin{cases}
        \Lambda_{\inf}^4\left( \frac{ 2 M_{\pl} \, n}{2  M_{\pl} \, n + 3 \sqrt{\alpha} \Lambda_{\fl}} \right)^{2n}\;,\quad\text{E-Model}\;,\\
        \Lambda_{\inf}^4\left(
\frac{
-3 + \sqrt{9 + \theta_4}
}{
3 + \sqrt{9 + \theta_4}
}
\right)^{2n}\;,\quad\text{T-Model}\;,\\
\Lambda_{\inf}^4\frac{
    \left( -\alpha \Lambda_{\fl}^2 + \left( \alpha^2 \Lambda_{\fl}^4 \Theta_2 \right)^{1/3} \right)^2}{
    \alpha^2 \Lambda_{\fl}^4 + 
    \left( \alpha^2 \Lambda_{\fl}^4 \Theta_2 \right)^{2/3}+
    \left( \alpha^5 \Lambda_{\fl}^{10} \Theta_2 \right)^{1/3}}\;,\quad\text{P-Model \,($n=1$)}\;,
    \end{cases}
\end{equation}
where
\begin{align}
    \theta_4 ~&:=~ \frac{24 M_{\pl}n\left( 4 M_{\pl}n + \sqrt{16 M_{\pl}^2 n^2 + 3 \alpha \Lambda_{\fl}^2} \right)}{\alpha \Lambda_{\fl}^2}\;,\\
    \Theta_2 ~&:=~ \left( \alpha \Lambda_{\fl}^2 + 6 M_{\pl} \left( 3 M_{\pl} + \sqrt{9 M_{\pl}^2 + \alpha \Lambda_{\fl}^2} \right) \right)\;.
\end{align}
Finally, we compute the number of e-folds $N_k$ between the instant at which a perturbation with wavenumber $k$ crosses the horizon and the end of inflation. Using \cref{eq:Nk}, we get
\begin{equation}
    \label{eq:Nk-appendix}
    N_k ~=~  \begin{cases}
        \frac{3\alpha \Lambda_{\fl}}{4 n M_{\pl}^2}\left( \ee^{\frac{\sqrt{\frac{2}{3\alpha}} \chi_k}{\Lambda_{\fl}}} \Lambda_{\fl}
- \ee^{\frac{\sqrt{\frac{2}{3 \alpha}} \chi_{\mathrm{end}}}{\Lambda f}}\Lambda_{\fl}
+\sqrt{\frac{2}{3\alpha}} (\chi_{\text{end}} - \chi k)
\right)\;\quad\text{E-Model}\;,\\
\frac{3 \alpha \Lambda_{\fl}^2}{16 M_{\pl}^2 n} \left(\cosh \left(2 \sqrt{\frac{2}{3\alpha}} \frac{\chi_k}{\Lambda_{\fl}}\right)- \cosh \left( 2 \sqrt{\frac{2}{3\alpha}} \frac{\chi_\mathrm{end}}{\Lambda_{\fl}} \right)\right)\;,\quad\text{T-Model}\;,\\
\frac{\left(\chi_k - \chi_{\mathrm{end}}\right)\left(\chi_{\mathrm{end}} + \chi_k\right)\left( 3 \alpha \Lambda_{\fl}^2 + \chi_{\mathrm{end}}^2 + \chi_k^2 \right)}{12 M_{\pl}^2 \alpha \Lambda_{\fl}^2}\;,\quad\text{P-Model \,($n=1$)}\;.
    \end{cases}
\end{equation}
\Cref{eq:epsilon-appendix,eq:eta-appendix,eq:ns-appendix,eq:r-appendix,eq:As-appendix,eq:chik-appendix,eq:Lambdainf-appendix,eq:chiend-appendix,eq:Vchiend,eq:Nk-appendix} can be used in both sides of \cref{eq:GammaKey}. Thus, for a given \ac{FN} model and fixing $\left(\Lambda_{\inf}, \alpha , n\right)$, we can solve for $n_s$. 

The running of the spectral index is defined as
\begin{equation}
\label{eq:alphasdlnk}
    \alpha_s ~=~ \frac{\dd n_s}{\dd \log k} \;.
\end{equation}
It can also be written as
\begin{equation}
\label{eq:alphasetaepsilonxi}
    \alpha_s ~=~ \left(16 \epsilon\eta  - 24\epsilon^2 - 2\xi^2 \right)\bigg|_{\chi = \chi_k}\;,
\end{equation}
where
\begin{equation}
    \xi^2 ~=~ M_{\pl}^2\frac{\partial_\chi V \left( \chi\right)}{V^2\left( \chi \right)}\,\partial_{\chi}^{(3)}V\left( \chi \right)\;.
\end{equation}
For the $\alpha-$attractor potentials, the expression for the running of the spectral index is given by
\begin{equation}
\label{eq:alphsSexplicit}
    \alpha_s ~=~ \begin{cases}
        \frac{M_{\pl}^4}{\alpha^2\Lambda_{\fl}^4}\frac{
32 \left( 8 - 13\, \ee^{\sqrt{\frac{2}{3\alpha}}\frac{\chi k}{\Lambda_\fl}} + \ee^{2\sqrt{\frac{2}{3\alpha}}\frac{ \chi k}{\Lambda _{\fl}}} \right)
}{
9 \left( 1 - \ee^{\sqrt{\frac{2}{3\alpha}}\frac{\chi_k}{\Lambda_{\fl}}} \right)^4
}\;,\quad\text{E-Model}\;,\\
\frac{256 M_{\pl}^4 }{9\alpha^2 \Lambda_{\fl}^4}\mathrm{csch}\left[ 2\sqrt{\frac{2}{3\alpha}}\frac{\chi_k}{\Lambda_{\fl}} \right]^4\\
\times\left( 19 - 28\cosh\left[ 2\sqrt{\frac{2}{3\alpha}}\frac{\chi_k}{\Lambda_{\fl}} \right] + \cosh\left[ 4\sqrt{\frac{2}{3\alpha}}\frac{\chi_k}{\Lambda_{\fl}}  \right] \right)\;,\quad\text{T-Model}\;,\\
864 M_{\pl}^4\alpha^2\Lambda_{\fl}^4 \frac{
\left(4\chi_k^4  - 3 \alpha^2\Lambda_{\fl}^4 - 18\alpha\Lambda_{\fl}^2\chi_k^2 \right)
}{
\left( 3\alpha\Lambda_{\fl}^2\chi_k + 2\chi_k^3 \right)^4
}\;,\quad\text{P-Model \,($n=1$)}\;.
    \end{cases}
\end{equation}

\section{Reheating Temperature for $n=1$ and $n=3$}
\label{app:SuppReheatingPlot}

\begin{figure}[H]
     \centering
     \begin{subfigure}[t!]{0.49\textwidth}
         \centering
         \includegraphics[width=\textwidth]{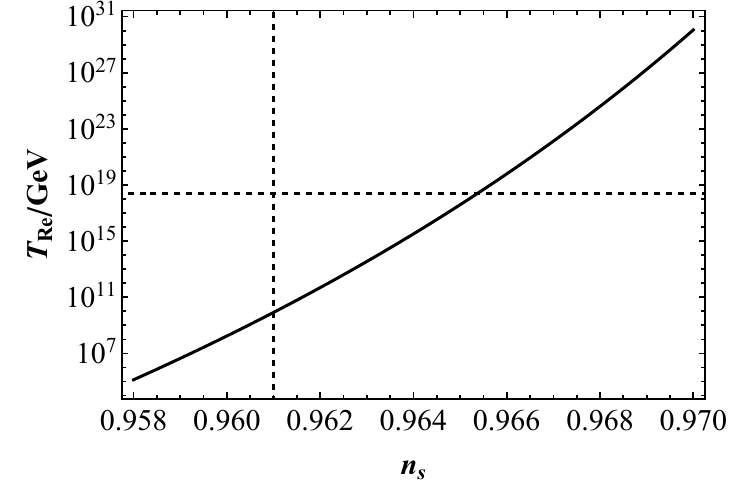}
         \caption{\it \label{fig:TreVns-n-1}$n=1$}
     \end{subfigure}
     \hfill
     \begin{subfigure}[t!]{0.49\textwidth}
         \centering
         \includegraphics[width=\textwidth]{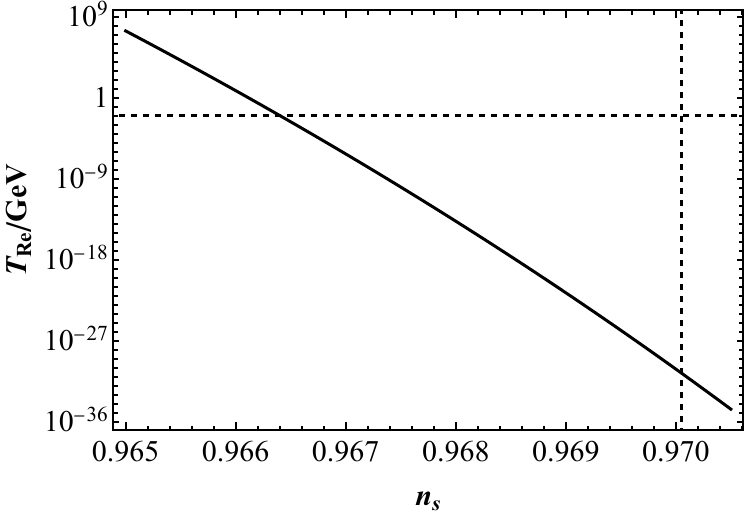}
         \caption{\it \label{fig:TreVns-n-3}$n=3$}
     \end{subfigure}
        \caption{\label{fig:TreVns}
       Plot of the reheating temperature $T_{\re}$ as a function of the spectral index $n_s$ for a classical non-vanishing mass ($n=1$) in the left panel and for a vanishing classical mass ($n=3$) in the right panel. Both panels are calculated for the $E$-model with $\alpha = 0.1$. The horizontal dashed line in the left panel indicates the reduced Planck scale $M_{\pl}$ while the horizontal dashed in the right represents the \ac{BBN} scale $1\,$MeV. In each panel, the vertical dotted line denotes the maximum decay width obtained for the models in \Cref{tab:models} 
        }.
\end{figure}

This Appendix contains the analysis for the reheating temperature $T_{\re}$ as a function of the spectral index $n_s$ for the $E-$model. In \Cref{fig:TreVns-n-1}, we show the reheating temperature $T_{\re}$ as a function of $n_s$ for $\alpha = 0.1$. We also include a vertical dotted line that indicates the maximum inflavon decay width, which corresponds to the highest reheating temperature obtained for the models listed in \Cref{tab:models}. It yields a corresponding reheating temperature of roughly $\sim 10^{11}\,$GeV. Thus, $T_{\re}$ is bigger than the  \ac{EW} symmetry crossover transition temperature, but well below the Planck scale denoted by the horizontal dotted line in \Cref{fig:TreVns-n-1}. Furthermore, the reheating temperature also remains smaller than $\Lambda_{\FN}$, which allows us to use the effective field theory for the \ac{FN} Lagrange density in \cref{eq:BeforeFlavorL}. We checked that these conditions: $150\,$GeV $<T_{\re} < \Lambda_{\FN}$ are satisfied for all the range of $\alpha$ values considered, and for the $T$ and $P$ models.

In \Cref{fig:TreVns-n-3}, we plot  the relation between the reheating temperature and the spectral index for $n=3$ and $\alpha = 0.1 $. Additionally, we indicate with a vertical dotted-line the maximum value of the reheating temperature from the different models in \Cref{tab:models}. We observe that the reheating temperature is much smaller than the \ac{BBN} temperature around $1\,$MeV. The energy transfer is significantly less efficient for $n=3$ than for $n=1$, which leads to this pronounced suppression in the reheating temperature. This outcome is not expected to change for different \ac{FN} charge assignments beyond those considered in this work. Moreover, note that a smaller contribution from the lepton sector, which can be satisfied by choosing a smaller $\Lambda_\nu$,  would further suppress the decay width. Thus, this decreases the reheating temperature $T_{\re}$ even more. We verified that, for the models in \Cref{tab:models}, the reheating temperature remains below the \ac{BBN} temperature around $1\,$MeV for all the $\alpha$ values considered in this work, as well as for $n = 3, 4, 5$ and for both the $T$ and $P$ models. Thus, an inflavon \ac{FN} model with $n\neq1$ cannot be physically realized within our framework.

\bibliography{biblio}
\bibliographystyle{NewArXiv}

\begin{acronym}
\acro{CMB}{Cosmic Microwave Background}
\acro{FN}{Froggatt-Nielsen}
\acro{DM}{dark matter}
\acro{SUSY}{supersymmetry}
\acro{VEV}{vacuum expectation value}
\acro{EFT}{effective field theory}
\acro{MFV}{minimal flavor violation}
\acro{SM}{Standard Model}
\acro{EW}{Electroweak}
\acro{NO}{Normal ordering}
\acro{IO}{Inverted ordering}
\acro{QCD}{Quantum chromodynamics}
\acro{FCNC}{Flavor-violating neutral current }
\acro{BBN}{Big Bang Nucleosynthesis}
\acro{CKM}{Cabibbo–Kobayashi–Maskawa}
\acro{PMNS}{Pontecorvo–Maki–Nakagawa–Sakata}
\acro{FRW}{Friedmann-Robertson-Walker}
\acro{CL}{confidence level}
\end{acronym}

\end{document}